\documentclass[aps,pre,reprint,superscriptaddress]{revtex4-1}

\usepackage{graphicx}
\usepackage{dcolumn}
\usepackage{bm}

\usepackage{amsmath}
\usepackage{color}

\begin{document}


\title{Relativistic two-wave resonant acceleration of electrons at large-amplitude standing whistler waves during laser-plasma interaction} 

\author{Takayoshi Sano}
\email{sano@ile.osaka-u.ac.jp}
\affiliation{Institute of Laser Engineering, Osaka University, Suita,
  Osaka 565-0871, Japan} 


\author{Shogo Isayama}
\author{Kenta Takahashi}
\author{Shuichi Matsukiyo}
\affiliation{Interdisciplinary Graduate School of Engineering Sciences,
Kyushu University, Kasuga, Fukuoka 816-8580, Japan}

\date{\today}

\begin{abstract}
The interaction between a thin foil target and a circularly polarized laser light injected along an external magnetic field is investigated numerically by particle-in-cell simulations.
A standing wave appears at the front surface of the target, overlapping the injected and partially reflected waves.
Hot electrons are efficiently generated at the standing wave due to the relativistic two-wave resonant acceleration if the magnetic field amplitude of the standing wave is larger than the ambient field.
A bifurcation occurs in the gyration motion of electrons, allowing all electrons with non-relativistic velocities to acquire relativistic energy through the cyclotron resonance.
The optimal conditions for the highest energy and the most significant fraction of hot electrons are derived precisely through a simple analysis of test-particle trajectories in the standing wave.
Since the number of hot electrons increases drastically by many orders of magnitude compared to the conventional unmagnetized cases, this acceleration could be a great advantage in laser-driven ion acceleration and its applications.
\end{abstract}


\maketitle


\section{Introduction}

Strong magnetic fields make considerable differences in the properties of relativistic laser-plasma interaction \cite{yang15,feng16,luan16,sano17,sano19,park19,sano20a,sano20c,hata21,karmakar21,weichman22}.
When laser injection along an external magnetic field is considered, electromagnetic waves propagate in plasmas as circularly polarized waves.
The whistler-mode wave appears when the cyclotron frequency of electrons $\omega_{ce}$ is larger than the laser frequency $\omega_0$.
The characteristics of the whistler wave have many attractive aspects for plasma heating and particle acceleration. 
The nature of right-hand circular polarization to the magnetic field has a strong affinity with the electron cyclotron resonance.
Since there is no cutoff density for propagation, it may interact directly with overdense plasmas and induce efficient energy transfer.
Furthermore, the whistler wave often plays a crucial role in various astrophysical phenomena such as planetary magnetospheres, stellar winds, and magnetars \cite{marsch06,turolla15,stenzel16}.
This paper focuses on how large-amplitude whistler waves efficiently generate relativistic hot electrons.

The propagation of whistler waves requires the existence of a supercritical magnetic field.
The critical strength is derived from the balance $\omega_{ce} = \omega_0$ and given by $B_c = m_e \omega_0 / e$, where $m_e$ is the electron mass and $e$ is the elementary charge.
For the case of high power lasers with the wavelength $\lambda_0 = 1$ $\mu$m, the critical strength corresponds to $B_c = 10 (\lambda_0 / 1 \mu{\rm m} )^{-1}$ kilo-Tesla.
Although quasi-static strong magnetic fields have been used in laser experiments for various purposes \cite{albertazzi14,arefiev16,matsuo19,law20,sano21b}, the current controllable field strength is limited and has yet to reach 10 kT.
However, the generation of axial magnetic fields in the mega-Tesla order has been proposed theoretically \cite{murakami20,shokov21,zosa22}, and thus, it could be possible that magnetic fields exceeding 10 kT would become available in laboratories.
Thus, we are trying to clarify the complex laser-plasma interaction under extreme unexplored conditions in advance.

Whistler waves can penetrate dense plasmas because they have no critical density.
The pulse length that can be propagated is constrained mainly by the stimulated Brillouin scattering, but it could survive at least for several hundred femtoseconds even in the cases of relativistic wave amplitudes \cite{sano20a,sano20c,hata21}.
We have reported that the whistler waves entering dense plasmas can directly heat the ions there \cite{sano19}.
In this process, the standing wave structure formed by two opposing whistler waves is essential.
Periodic density bunches of electrons are formed in the standing wave, and a large-amplitude longitudinal electric field is excited. 
This electric field drives a compressive wave of ions, which eventually thermalizes due to plasma instabilities, producing a high-temperature ion plasma.
Since the injected wave energy is converted to ions without going through electrons, this heating process could be utilized as an efficient alternative scheme for fast ignition laser fusion \cite{sano20a}.

This paper sheds light on electron acceleration by standing whistler waves in underdense plasma.
The cyclotron resonance energizes electrons to relativistic speeds.
In previous studies, efficient energy conversion from lasers to electrons occurs when a magnetic field of appropriate strength is applied \cite{sano17,sano19}. 
However, the physical mechanism still needed to be fully understood, although numerical simulations suggested that the cyclotron resonance of relativistic electrons was the primary factor.
Therefore, we revisit the issue of electron acceleration by whistler waves.
This work includes the effects of particle-particle collisions and preplasma, which were not considered in the previous analysis.
It has been pointed out that the standing wave structure plays a key role in electron acceleration \cite{matsukiyo09,lee13,isayama23}.
The acceleration is caused by simultaneous relativistic resonance between a particle and two different waves, which is referred to in this paper as the "relativistic two-wave resonant acceleration".
Then, we investigate the motion of electrons in standing whistler waves using both numerical simulations and an analytical approach to particle trajectories.

When laser light irradiates a solid target, injected and reflected components form a standing wave at the front side of the target.
Then, underdense preplasma will interact with the standing wave if it exists.
Previous works have examined cases without an external magnetic field and revealed that a stochastic process in the standing wave efficiently generates hot electrons
\cite{sheng04,bochkarev19,huller19,weichman20}.
However, in our cases with a strong magnetic field, electron acceleration is caused by the cyclotron resonance.
The fraction of relativistically accelerated electrons shows significant enhancement compared to unmagnetized situations.

Higher temperatures and higher number densities of relativistic hot electrons are of great advantage for laser-driven ion acceleration.
The hot electrons generated by laser irradiation to a thin foil target form a strong sheath electric field on the backside of the target.
In recent years, ion beams accelerated by this sheath field have attracted much attention for plasma diagnostics and medical cancer therapy \cite{daido12}.
For medical applications, the energy of laser-driven protons and heavy ions is not yet sufficient, and then the improvement of the maximum ion energy is an urgent issue.
Then, it must be meaningful to investigate quantitatively how much improvement can be expected by combining a strong magnetic field with an intense laser.

Here, we investigate the generation process of hot electrons during the interaction between a dense plasma target and a large-amplitude electromagnetic wave traveling along an external magnetic field.
The outline of this paper is as follows.  
In Sec.~\ref{sec2}, the numerical setup for particle-in-cell (PIC) simulations is described.  
Various simulation results are shown in Sec.~\ref{sec3} to reveal the electron acceleration mechanism under a strong magnetic field.
We scrutinize parameter dependences on the magnetic field strength and the laser intensity by a series of one-dimensional simulations. 
Then, the physical mechanism is identified by an analytical approach using test-particle trajectories in standing electromagnetic waves.
The application to laser-driven ion acceleration is discussed in Sec.~\ref{sec4}. 
The robustness of our acceleration mechanism is also mentioned in the discussion.
Finally, the conclusions are summarized in Sec.~\ref{sec5}.

\section{Numerical Setup and Method \label{sec2}}

We consider a foil target in the vacuum irradiated by laser light from one side. 
In general, if there is a discontinuity in the plasma density, a fraction of a traveling electromagnetic wave should be reflected at the boundary and form a standing wave by overlapping the incident and reflected waves.
Then, a standing wave appears at the front surface of the target.
The electron motions in the standing wave are solved numerically and analytically in our analysis.

Another crucial element for the electron acceleration is an external magnetic field $B_{\rm ext}$ with sufficient strength.
We are particularly interested in the cases where the direction of the magnetic field is set parallel to the laser injection.
As for the strength, the electron cyclotron frequency $\omega_{ce} = e B_{\rm ext} / m_e$ is mostly assumed to be more than the laser frequency $\omega_0$, but that of ion $\omega_{ci}$ is much less than $\omega_0$.
Thus the electromagnetic waves we focused on correspond to the whistler-mode wave.

The laser light should be injected from the vacuum region to the target foil.
The laser wavelength in the vacuum is $\lambda_0 = 2 \pi c / \omega_0$, where $c$ is the speed of light. 
The electric-field amplitude of the laser $E_0$ is denoted by the normalized vector potential $a_0 = e E_0 / (m_e c \omega_0)$.
The pulse duration $\tau_0$ is defined by the full-width at half maximum of Gaussian-shaped envelope.
The polarity of the laser is also one of the control parameters of the electron acceleration.
Right-hand circularly polarized (RCP) waves to the magnetic field direction are investigated primarily because of a smooth mode conversion to whistler waves in the plasma.

As the target material, diamond is adopted. 
Although any material can be used, we have chosen it thinking of its application to carbon-ion acceleration \cite{daido12}.
The electron density in the target is then set to $\widetilde{n}_e \equiv n_e / n_c =603$.
Equating the plasma and laser frequencies, $\omega_{pe} = \omega_0$, the critical density is given by $n_c = \epsilon_0 m_e \omega_0^2 / e^2$, where $\epsilon_0$ is the vacuum permittivity.
The target thickness is typically $d = \lambda_0$.
As usual, the preplasma with an exponential distribution is considered at the front side of the target.
The scale length is assumed to be $\lambda_0$ with the spread width of $5 \lambda_0$.
The maximum density of the preplasma is $10 n_c$ adjacent to the target surface, which is less than the constant density in the target.
For simplicity, we assume that the carbon ions are fully ionized as C$^{6+}$ from the beginning. 
However, this assumption has little influence on our conclusions since the interaction with a relativistic intensity laser would quickly result in the field ionization inside the target due to the intrusion of whistler waves.

A uniform magnetic field is applied perpendicular to the target surface.
The normalized strength $\widetilde{B}_{\rm ext} \equiv B_{\rm ext}/B_c$ ($>0$) is used in the following analysis.
Then, this system is characterized predominantly by three non-dimensional parameters; $\widetilde{n}_e$, $\widetilde{B}_{\rm ext}$, and $a_0$.

The incident wave reaching the target surface must be partially reflected if there is a density jump.
The whistler-mode wave can travel in any density plasmas if the magnetic field strength is more than $B_c$. 
Thus, an injected RCP wave from the vacuum will be converted to a whistler wave with the same frequency $\omega_0$ at the target surface. 
The transmittance of the electric field $E_{\rm tra} / E_0 = 2 / (N + 1)$ is characterized by the refractive index 
$N$ of the whistler wave for the target density \cite{chen84},
\begin{equation}
N = \left( 1 - \frac{\widetilde{n}_e}{1 - \widetilde{B}_{\rm ext}} \right)^{1/2} \;.
\label{eq:n}
\end{equation}
Then, the reflected component is always non-zero at the target surface and contributes to the standing wave formation.

The wave-plasma interaction is solved by a PIC scheme, PICLS \cite{sentoku08}, including the Coulomb collisions.   
For one-dimensional configuration, we set the $x$-coordinate as the direction of the external magnetic field.
The origin $x = 0$ is defined at the front surface of the target.
The RCP laser light is injected from the minus-side boundary of the computational domain and propagates in the forward direction of the $x$-axis.
The main pulse hits the target from $t = 0$ to $\tau_0$, which defines the time reference point in our simulations.
The vacuum regions on both sides of the target are kept reasonably vast, more than $50 \lambda_0$.
The escape boundary conditions for waves and particles are adopted for both boundaries.

The spatial and temporal resolution is $\varDelta x = c \varDelta t = \lambda_0 / 10^3$.
The particle numbers for ions and electrons are initially 100 and 600 per grid cell, respectively.
The particles in the PIC simulations are so-called superparticles, which are aggregates of multiple particles.
In our scheme, the initial density distribution is expressed by changing the weight of the superparticle.
The particle number per cell is the same anywhere in the target and preplasma.
Nevertheless, the number weight of each superparticle is different according to the plasma density at each location.

In strongly magnetized plasmas, the time resolution $\varDelta t$ should be shorter than the electron gyration time, as well as the laser period.
Otherwise, the numerical heating breaks the energy conservation.
It would be better to resolve the wavelength by at least a few hundred grid cells in order to capture the propagation of whistler waves and the evolution of standing waves accurately. 
These conditions are satisfied in all simulations shown in this paper.
The convergence test has confirmed that the conclusions of our analysis are unaffected by the numerical resolution.

The list of performed parameters of main models and the obtained results for characteristic quantities are summarized in three tables in Appendix~\ref{app3}.
It will serve as a reference for the discussions of numerical results in the following sections.

Hereafter, the length and time are normalized by the laser wavelength $\lambda_0$ and period $t_0 = 2 \pi / \omega_0$, which will be written by the variables with a tilde, $\widetilde{l} = l / \lambda_0$ and $\widetilde{t} = t / t_0$.

\section{Numerical Results \label{sec3}}

\subsection{Spectral features of hot electrons}

\begin{figure*}
\includegraphics[scale=0.85,clip]{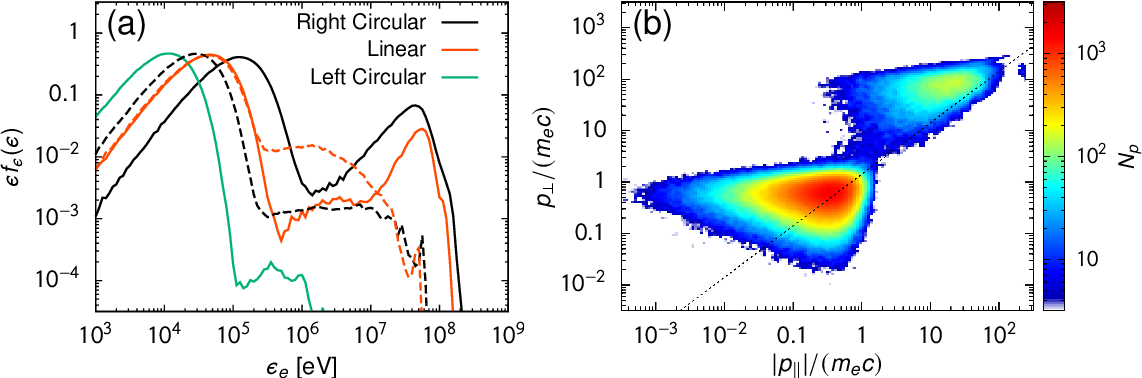}%
\caption{
(a) 
Electron energy spectra after the interaction of a thin foil target and a laser light propagating along an external magnetic field.
The target material is solid carbon with the density of $n_e / n_c = 603$ and the foil thickness is $d / \lambda_0 = 1$.
The polarization of the laser is right-hand circular (black solid; run01), linear (red solid; run03), and left-hand circular (green; run02).
The laser amplitude is $a_0 = 30$ for circularly polarized waves and $a_0 = 42$ for linearly polarized one, and the pulse duration is $\tau_0 / t_0 = 10$ for all the cases.
The external magnetic field is assumed to be $B_{\rm ext} / B_c = 30$.
For comparison, corresponding unmagnetized cases (run05 and run06) are shown by the dashed curves with the same colors.
All the spectra are measured at $t / t_0 = 71$.
(b)
Particle number distribution of electrons $N_p$ in the momentum phase space of $|p_{\parallel}|$ and $p_{\perp}$ for the fiducial case of right-hand circularly polarized laser, which is depicted by the black solid curve in (a).
The dotted line denotes the isotropic condition $p_{\perp} = 2 |p_{\parallel}|$.
\label{fig1}}
\end{figure*}

The generation process of hot electrons via laser-plasma interaction is significantly affected by the polarization of the incident laser and the external magnetic field.
First, the fundamental features are demonstrated through several runs with fiducial parameters.
Figure~\ref{fig1}(a) shows the electron energy spectra taken at $\widetilde{t} = 71$ after a sufficient time has passed since the interaction.
The model parameters of the fiducial run are $\widetilde{B}_{\rm ext} = 30$ and $a_0 = 30$ for RCP wave injection.
The pulse duration is $\widetilde{\tau}_0 = 10$ and the thickness of the carbon target is $\widetilde{d} = 1$.
The laser incident angle to the magnetic field is assumed to be $\theta = 0$ as default.

By choosing the wavelength $\lambda_0 = 0.8$ $\mu$m, the critical values are evaluated as $n_c = 1.72 \times 10^{21}$ cm$^{-3}$ and $B_c = 13.4$ kT. 
Then, the dimensional parameters of the fiducial run correspond to $n_{e} = 1.05 \times 10^{24}$ cm$^{-3}$, $B_{\rm ext} = 402$ kT, $I_0 = 3.85 \times 10^{21}$ W/cm$^{2}$, $\tau_0 = 26.7$ fs, and $d = 0.8$ $\mu$m.
The mass density of the target is $\rho = 3.51$ g/cm$^{-3}$, which is equivalent to diamond.
The laser conditions are determined based on the typical quantities of TW-class femtosecond lasers.

When the RCP laser is irradiated, the energy distribution of electrons has two distinct peaks around 100 keV and 30 MeV for the bulk and hot components, respectively [see the black solid curve in Fig.~\ref{fig1}(a)].
The number fraction of hot electrons is more than 10\% for this case.
Compared with the spectrum of the left-hand circularly polarized (LCP) case (green curve), it is evident that the RCP wave makes a remarkable contribution to the generation of hot electrons.
Even for the RCP laser, the second peak of hot electrons disappears in the absence of the external magnetic field (black dashed curve).
The hot electron fraction drops below 1\%, and thus the magnetic field makes a notable difference in the production rate by more than an order of magnitude.
Therefore, the RCP wave and the strong magnetic field must be indispensable elements for the electron acceleration mechanism.

The results for the linearly polarized (LP) laser are also shown in Fig.~\ref{fig1}(a) by red curves.
The pulse duration is the same as the CP cases, but the amplitude is set to $a_0 = 42$ to equalize the input wave energy.
The electron features in the spectra are very similar to those in the RCP runs.
The second peak of hot electrons is also visible in the LP run, and the peak energy is about the same as in the RCP case. 
However, the number of hot electrons is considerably reduced by a factor of one-third.
The LP wave can be regarded as a superposition of the RCP and LCP waves of half amplitude. 
The RCP component within the LP laser must be responsible for the hot electron production.

The momentum distribution of accelerated hot electrons is found to be anisotropic.
The phase diagram of $|p_{\perp}|$ and $p_{\parallel}$ is depicted in Fig.~\ref{fig1}(b) for the fiducial RCP run with the magnetic field.
Here, the momentums are defined as $\bm{p} = \gamma m_e \bm{v}$, $p_{\parallel} = p_x$, and $p_{\perp} = (p_y^2 + p_z^2)^{1/2}$.
Two components of the bulk and hot electrons are separated clearly in this figure.
The bulk component is subject to the dotted line of $p_{\perp} = 2 | p_{\parallel} |$, indicating that it is isotropic.
The hot component, on the other hand, exhibits $p_{\perp} \gtrsim 2 | p_{\parallel}|$, so that the gyration velocity is dominant.
It infers that the cyclotron resonance would be deeply involved in this acceleration process. 

\subsection{Acceleration site in the standing wave}

\begin{figure*}
\includegraphics[scale=0.85,clip]{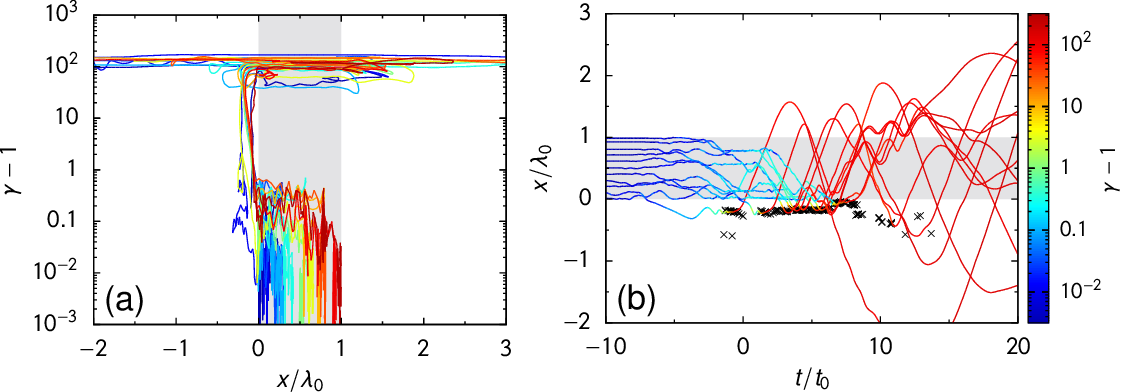}%
\caption{
(a) 
Trajectories of hot electrons shown in the diagram of the position $x$ and kinetic energy $\gamma -1$.
Eleven particles are selected among the hot electrons that are eventually accelerated to over $\gamma = 100$.
The front surface of the target is $x = 0$, and the gray-filled area indicates the target thickness.
The initial locations of these accelerated particles are inside the target, and each trajectory curve is drawn by different colors.
(b)
Electron trajectories in the position-time plane for the same 11 hot electrons in (a).
The line color denotes the kinetic energy of the electron at each time.
The gray area shows the original target location.
The cross marks mean the place where the hot electrons are accelerated and exceed the relativistic energy defined as $\gamma = 30$.
\label{fig2}}
\end{figure*}

Before proceeding to the theoretical modeling of the acceleration mechanism, we will check some more properties of hot electrons that can be revealed from our PIC simulations.

The $x$ position where all the relativistic electrons are accelerated is turned out to be outside of the target surface.
The behavior of hot electrons in the position-energy diagram is shown in Fig.~\ref{fig2}(a).
We picked up 11 particles that finally gained energy up to $\gamma > 100$ starting from the inside of the dense target.
The trajectories of the electrons with different initial positions are displayed in different colors.
All the particles are regularly accelerated near the target surface from a non-relativistic velocity ($\gamma < 1$) to a relativistic velocity ($\gamma \sim 100$).
Interestingly, as the $x$-coordinate is constant during the acceleration, a sudden increase in energy by more than two orders of magnitude occurs at a fixed location.
In other words, this rapid acceleration is associated with only the perpendicular velocity to the external magnetic field.

The longitudinal velocity increases after the electron energy reaches around $\gamma \sim 100$.
The recirculation motion back and forth between the front and rear sides of the target is repeated with relativistic speed.
In Fig.~\ref{fig2}(b), the trajectories of the selected 11 particles are illustrated in the time-position plane, where the line color indicates the electron kinetic energy.
All the accelerated particles migrate once toward the front side of the target and then acquire the energy just outside the target.
The acceleration time as indicated by the color change from blue to red is much shorter than the pulse duration of $\widetilde{\tau}_0 = 10$ and comparable to the laser oscillation time.

The locations and timings where the energy of the accelerated electrons exceeds $\gamma = 30$ are also plotted in Fig.~\ref{fig2}(b).
Note that these cross marks contain information about all the hot electrons, including not only the selected ones.
The term when the acceleration takes place coincides to the laser irradiation time ($0 < \widetilde{t} < 10$), and the position is about at $\widetilde{x} \sim - 0.25$.
The acceleration site is slightly outside the target and not inside the target at all. 
Besides, all electrons in the preplasma are accelerated to relativistic velocity in situ. 
These facts suggest that the standing waves generated outside the target would play an important role in the relativistic electron acceleration.

\begin{figure*}
\includegraphics[scale=0.85,clip]{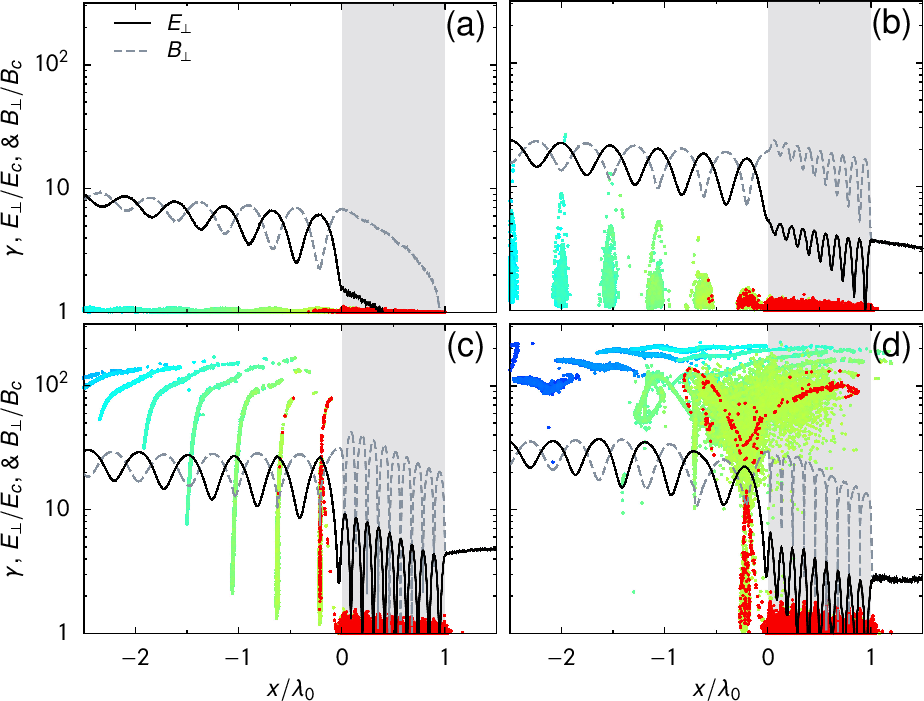}%
\caption{
Time variation of the envelope of the electromagnetic fields, $E_{\perp}$ (black solid) and $B_{\perp}$ (gray dashed), and the electron energy (color) as a function of the position $x$.
The snapshots are taken at (a) $t / t_0 = -7$, (b) $-3$, (c) $-1$, and (d) 1 in the fiducial run of right-hand circularly polarized laser ($a_0 = 30$ and $B_{\rm ext} / B_c = 30$).
The color of each electron indicates the weight of the number of particles.
Particles initially inside the target are shown in a reddish color.
On the other hand, the particles in a bluish color are started from the preplasma region, which contain a relatively fewer number of particles as superparticles in the PIC simulation.
The gray area stands for the original target location.
\label{fig3}}
\end{figure*}

Let us check the electromagnetic fields at the acceleration site.
Figure~\ref{fig3} shows the time evolution of the tangential components, $E_{\perp} = (E_y^2 + E_z^2)^{1/2}$ and $B_{\perp} = (B_y^2 + B_z^2)^{1/2}$, near the target.
Each snapshot is taken at $\widetilde{t} = -7$, $-3$, $-1$, and 1, which is about the time when the main pulse arrives the target surface.
The gray area indicates the initial position of the carbon foil.
At $\widetilde{t} = -7$, the tip of the laser pulse has already traveled to the vicinity of the target rear side.
The magnetic field amplitude is dominant inside the target compared to the electric field because it propagates as a whistler wave.
Some fraction of the incident wave is reflected at the target front surface.
A standing wave is established there, and the amplitude is almost stationary from $\widetilde{t} = -3$ to 1.
For the case of CP waves, the amplitude of the standing wave at each position is constant in time.
The presence of the standing wave is recognized by nodes and antinodes in the distribution of the electromagnetic fields.

There is a strong correlation between the periodic structure of the electromagnetic field and the electron acceleration point.
The acceleration occurs at approximately equal intervals, and all the sites are precisely at the trough of the magnetic field in the standing wave.
Figure~\ref{fig3} also shows the information on the spatial distribution and energy of electrons by colored dots, where the color represents the number weight of superparticles in the PIC simulation.
Particles initially inside the target are shown in red, and electrons that started in the preplasma are green or blue.
Before the main pulse arrives, the electrons are all at $\gamma = 1$ because their energy is non-relativistic.
However, as the amplitude of the standing wave increases after $\widetilde{t} \gtrsim -3$, the electrons begin to accelerate in the preplasma region.
The electrons gradually gain energy as they gather in the trough of the magnetic field.
Finally, all the electrons are launched simultaneously like a line of fountains to the relativistic velocity of $\gamma \gtrsim 100$.
At the time $\widetilde{t} = 1$, almost all electrons in the preplasma are transferred to the relativistic speed.
Then, the generated cavitation structure in the electron density has quickly disappeared.

When we look at the particles inside the target (red dots), the accelerated electrons go through the closest trough of the magnetic field at $\widetilde{x} \sim - 0.25$ without exception.
The energies of the accelerated electrons are nearly identical, which is about $\gamma \sim 100$ in this case.
Due to the low density outside the target, the wavelength of the electromagnetic wave remains almost $\lambda_0$.
The interval between the standing wave nodes is half of $\lambda_0$.
Then, the trough (crest) of the magnetic (electric) field will locate at $\widetilde{x} = - 1/4$, $-3/4$, $-5/4$, $\cdots$, which are matching with the accelerateion site.
The standing wave persists for the laser irradiation time, during which the electrons that stray near the target surface are successively elevated to a relativistic speed.
As a result of these processes, an apparent dichotomy between the bulk and relativistic hot components emerges in the electron energy distribution.

Even when we use a bare target without the preplasma (e.g., run04 and run07), the behavior of hot electrons is unchanged qualitatively and even quantitatively.
However, this picture is entirely different from the cases without the initial magnetic field.
The hot electron fraction is drastically reduced because the relativistic electrons are generated only from the preplasma in that case.
The energy conversion from the laser to the electrons is diminished from 49\% to 0.9\% by simply removing the magnetic field (see run01 and run05 in Table~\ref{tab1}).
The collisional effects have little influence on the hot electron generation, although the bulk component takes longer to reach the thermal equilibrium.

The whistler wave propagating inside the target is partially transmitted at the rear boundary, and the rest is reflected.
Therefore, another standing wave is created within the target.
However, the same acceleration does not occur within the dense target, which may be because the amplitude of the electric field is weaker.
Furthermore, the plasma frequency there is much larger than $\omega_0$, and the collective movement of electrons cannot be ignored.
Although we concentrate our discussion on the diamond target in this paper, the dependence of the target density will be addressed in the subsequent analysis.

\subsection{Dependence on the external magnetic field}

\begin{figure*}
\includegraphics[scale=0.85,clip]{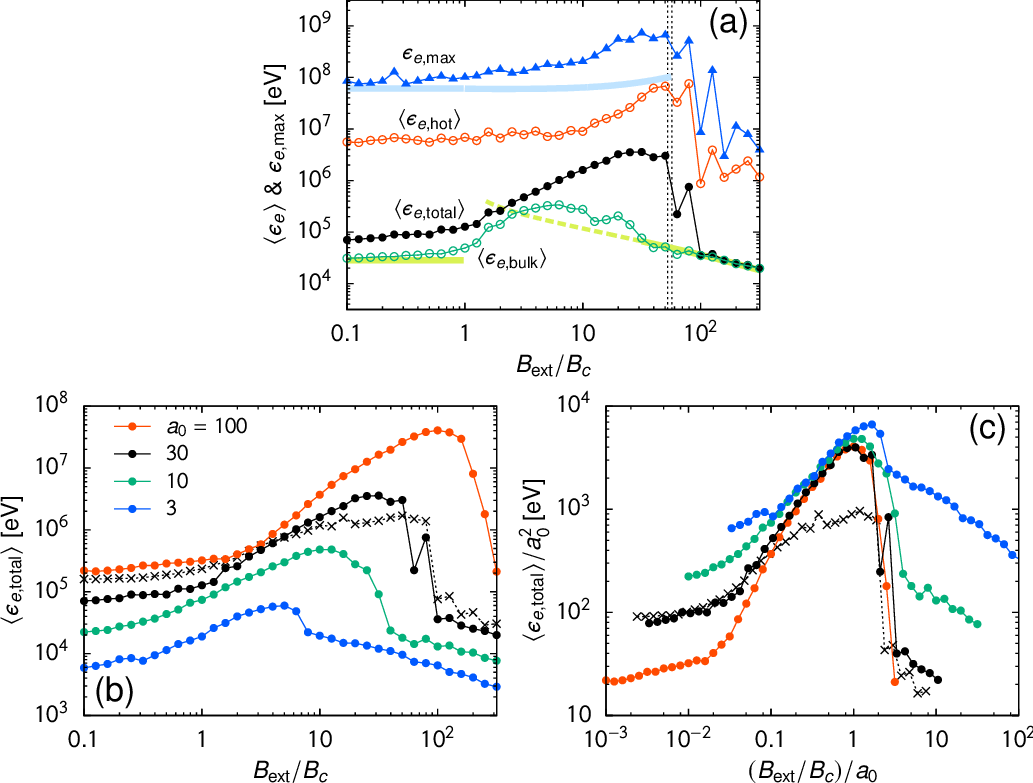}%
\caption{
(a)
The dependence of the maximum and average energies of electrons on the strength of an external magnetic field $B_{\rm ext} / B_c$.
The maximum energy $\epsilon_{e.\max}$ is shown by blue triangles, the average energies of total electrons $\langle \epsilon_{e,{\rm total}} \rangle$, of hot electrons $\langle \epsilon_{e,{\rm hot}} \rangle$, and of bulk electrons $\langle \epsilon_{e,{\rm bulk}} \rangle$ are plotted by black filled circles, red open circles, and green open circles, respectively.
The theoretical predictions of the maximum electron energy and the
bulk electron temperature are also denoted by the thick curves of
light-blue and light-green, which are discussed later on in Sec.~\ref{sec3}.D.
The vertical dotted lines at $B_{\rm ext} / B_c = 56.9$ and 52.4 indicate the theoretical boundaries given by Eqs.~(\ref{eq:a1}) and (\ref{eq:a2}), respectively.
(b)
The dependence of the average electron energy on the injected laser intensity $a_0$.
A wide range of the laser amplitude is examined; $a_0 = 100$ (red), 30 (black), 10 (green), and 3 (blue).
The other model parameters are identical to the fiducial run.
The polarization of the laser is assumed to be right-hand circular except for the runs shown by the cross marks, which are the cases of linearly polarized laser with the amplitude of $a_0 = 42$.
(c) The same data as in (b) but plotted with a different normalization.
The horizontal and vertical axes are divided by $a_0$ and $a_0^2$, respectively.
Then the acceleration feature at $1 \lesssim B_{\rm ext}/B_c \lesssim a_0$ is overlapped nicely for all the runs.
The energy peak is always achieved when $B_{\rm ext}/B_c \sim a_0$.
\label{fig4}}
\end{figure*}

It would be meaningful to clarify the dependence of the external magnetic field on this acceleration mechanism.
Then, we will examine how the electron energy distribution is affected by the strength of the magnetic field.
Figure~\ref{fig4}(a) shows the average and maximum energies of electrons as a function of $\widetilde{B}_{\rm ext}$.
All the parameters except for $\widetilde{B}_{\rm ext}$ are the same as in the fiducial run, and then the amplitude of incident RCP wave is $a_0 = 30$.
The average energy is measured not only as of all electrons, $\langle \epsilon_{e,{\rm total}} \rangle$, but also as of the bulk and hot components separately, $\langle \epsilon_{e,{\rm bulk}} \rangle$ and $\langle \epsilon_{e,{\rm hot}} \rangle$ (see Appendix~\ref{app1} for the evaluation method).

Focusing on the average energy of total electrons $\langle \epsilon_{e,{\rm total}}\rangle$, it is almost constant when the magnetic field strength is in the range of $\widetilde{B}_{\rm ext}\lesssim 1$.
However, it increases systematically as $\widetilde{B}_{\rm ext}$ exceeds unity, and takes a peak value around $\widetilde{B}_{\rm ext}\sim 30$.
In this moderate strength regime, the bulk energy $\langle \epsilon_{e,{\rm bulk}} \rangle$ increases by an order of magnitude around $\widetilde{B}_{\rm ext} \sim 3$ and then begins to decrease.
Whereas the hot component $\langle \epsilon_{e,{\rm hot}}\rangle$ increases monotonically with $\widetilde{B}_{\rm ext}$.
This trend changes abruptly when the external magnetic field becomes stronger than several tens of $B_c$, where the energy absorption rate of electrons drops sharply.
If the magnetic field is too strong, very few hot electrons are generated, as anticipated from the relation of $\langle \epsilon_{e,{\rm total}} \rangle \approx \langle \epsilon_{e,{\rm bulk}} \rangle$.
The variation of the maximum energy $\epsilon_{e,\max}$ behaves similarly to the average energy of hot electrons.
The maximum energy near the peak is about 1 GeV, and the average of hot electrons reaches 100 MeV.
Thus, the valid range in which electrons can be accelerated effectively is roughly $1 \lesssim \widetilde{B}_{\rm ext} \lesssim 30$ for this particular case.

The enhancement of the hot electron fraction is also dependent on the laser intensity.
In Fig.~\ref{fig4}(b), the average energies of electrons $\langle \epsilon_{e, {\rm total}} \rangle$ are evaluated when the laser amplitude has changed to $a_0 = 100$, 30, 10, and 3.
The higher the laser intensity, the higher the average energy of electrons.
In all cases, the electron energy rises when the magnetic field strength is larger than $B_c$.
As seen in the related work \cite{sano17}, the enhancement range in $\widetilde{B}_{\rm ext}$ for the energy conversion rate to electrons extends with increasing the laser intensity $a_0$.
The magnetic field strength at which the average energy is maximized is $\widetilde{B}_{\rm ext}\sim a_0$ approximately, and the range of efficient electron acceleration can be commonly expressed as $1 \lesssim \widetilde{B}_{\rm ext} \lesssim a_0$.
The average energy of total electrons is proportional to $a_0^2$ [see Fig.~\ref{fig4}(c)] and reaches 30 MeV if $a_0 = 100$.
Such enhancement gives a considerable difference of more than two orders of magnitude compared to the case without the initial magnetic field.

For comparison, the outcomes of LP light with $a_0 = 42$ are also shown by the cross marks in Fig.~\ref{fig4}(b).
The trend of the average energy is quite similar to the RCP cases.
In the limit of the weaker magnetic field, LP light exhibits a slight advantage in the efficiency of electron heating, but the difference is insignificant when $\widetilde{B}_{\rm ext}$ is greater than unity.
This tendency is consistent with the spectral features shown in Fig.~\ref{fig1}(a).

\begin{figure}
\includegraphics[scale=0.85,clip]{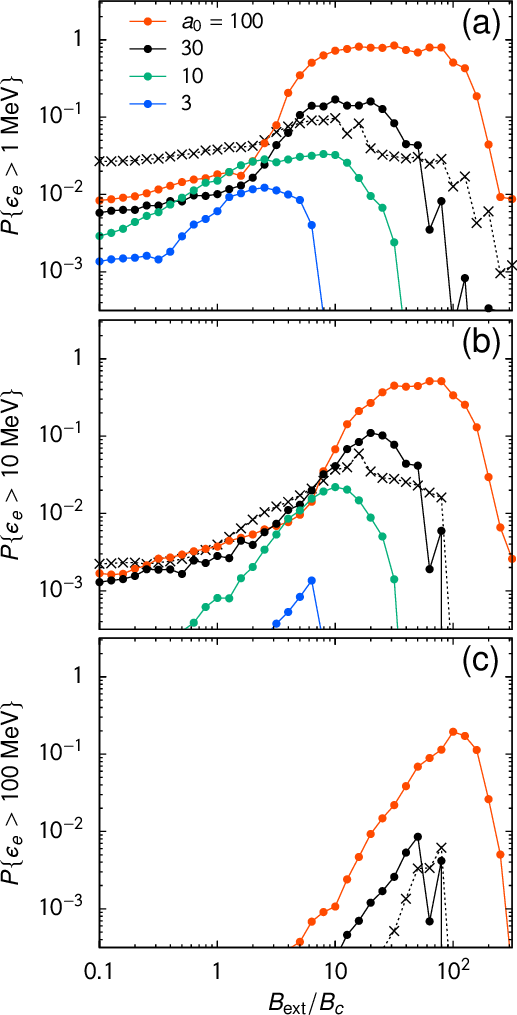}%
\caption{
The number fraction of hot electrons over the threshold energies of (a) $\epsilon_{\rm thr} = 1$ MeV, (b) 10 MeV, and (c) 100 MeV.
The meanings of the marks are identical to those of Fig.~\ref{fig4}(b).
\label{fig5}}
\end{figure}

The most striking feature of this electron acceleration is that it can enlarge the number of hot electrons.
Here, we denote the fraction of electrons with the energy above a certain threshold as $P \{\epsilon_e > \epsilon_{\rm thr}\}$.
Figure~\ref{fig5} illustrates the number fractions when the threshold is set to $\epsilon_{\rm thr} = 1$, 10, and 100 MeV.
As with the average energy, the number of hot electrons is boosted by the stronger magnetic field and larger amplitude of electromagnetic waves.

In order to organize the characteristics of hot electrons, we will use the results of the $a_0 = 30$ runs (the black circles in Fig.~\ref{fig5}) as an example.
In the range of $1 \lesssim \widetilde{B}_{\rm ext} \lesssim 30$, electrons over 1 MeV are enhanced by a factor of 10 or more compared to the unmagnetized limit.
Since the maximum energy is positively correlated with the magnetic field strength, a significant number of electrons breaks through 100 MeV at $\widetilde{B}_{\rm ext} \sim 30$.
The electron fractions above 10 MeV and 100 MeV are surprisingly large, exceeding 0.1 and 0.01, respectively.

If a more intense laser is available, like $a_0 = 100$, all electrons inside the target could be accelerated entirely to over 1 MeV.
Furthermore, the fraction of ultrarelativistic electrons with energies above 100 MeV reaches 20\% when $\widetilde{B}_{\rm ext} = a_0$ (see run10 in Table~\ref{tab2}).
Such fast electrons create a large sheath field on the target surface, which may increase the maximum energy and efficiency of ion acceleration. 
This curious application will be discussed later in Sec.~\ref{sec4}.
A similar property of the high energy conversion rate to electrons has been observed when the laser intensity becomes lower with keeping the relation $\widetilde{B}_{\rm ext} = a_0$ (e.g., run08 and run09). 
However, the merit of this acceleration mechanism cannot be exploited because the number of relativistic electrons is far from the dominant. 
In short, an RCP laser of sufficiently relativistic intensity and a coherent external magnetic field stronger than the critical strength is essential for this electron acceleration mechanism to become prominent.

\subsection{Analytical approach for the acceleration physics \label{sec3d}}

The electron acceleration phenomenon seen in the laser-plasma interaction under a strong magnetic field has many similarities to the acceleration mechanism discovered in another context.
Electron acceleration in the standing Alfv{\'e}n wave has been identified for the first time in one-dimensional turbulence of electron-positron plasmas \cite{matsukiyo09}.
If the wave amplitude exceeds a critical value, bifurcations occurs in the electron motions, allowing the conversion from non-relativistic to relativistic velocities without the injection problem \cite{isayama23}.
The crucial process is the cyclotron resonance of electrons with two CP waves that constitute a standing wave.
Here, we will extend and generalize their analysis of particle trajectories and compare it with the characteristics of our simulation results.

The equations of motion in the electromagnetic fields of standing wave determine the time evolution of the normalized electron momentum $\widetilde{\bm{p}} = \gamma \bm{v} / c$.
Because the electron acceleration occurs at a fixed location, those fundamental equations can be simplified by assuming that there is no longitudinal component $\widetilde{p}_{\parallel} = 0$.
The gyration motion of electrons at the acceleration point, or the trough of the magnetic-field envelope, is then described in a Hamiltonian form \cite{matsukiyo09}.
The conserved quantity, $H (\chi, \psi)$, is a function of the square of the electron momentum, $\chi = \widetilde{p}_{\perp}^2$, and the phase of the electron motion relative to the standing wave, $\psi$,
\begin{equation}
H (\chi, \psi) = A \chi^{1/2} \sin \psi - B (\chi + 1)^{1/2} + \chi \;,
\label{eq:ham}
\end{equation}
where $A = 2 (1+R) a_0$ and $B = 2 \widetilde{B}_{\rm ext}$  (see Appendix~\ref{app2} for the derivation).
The standing wave is formed from the incident laser and the reflected component, whose amplitude is $R a_0$.
The reflectivity $R$ is estimated using the refractive index as $R = (N - 1)/(N + 1)$, which is then given by a function of the electron density $\widetilde{n}_e$ and the magnetic field strength $\widetilde{B}_{\rm ext}$. 
In other words, coefficient $A$ includes a novel extension to account for differences in the amplitude of the opposing waves.
We will explain how the electron orbital feature described by Eq.~(\ref{eq:ham}) undergoes bifurcations depending on the amplitude of the standing wave, $(1+R)a_0 = A/2$, and the intensity of the external magnetic field, $\widetilde{B}_{\rm ext} = B/2$.

\begin{figure*}
\includegraphics[scale=0.85,clip]{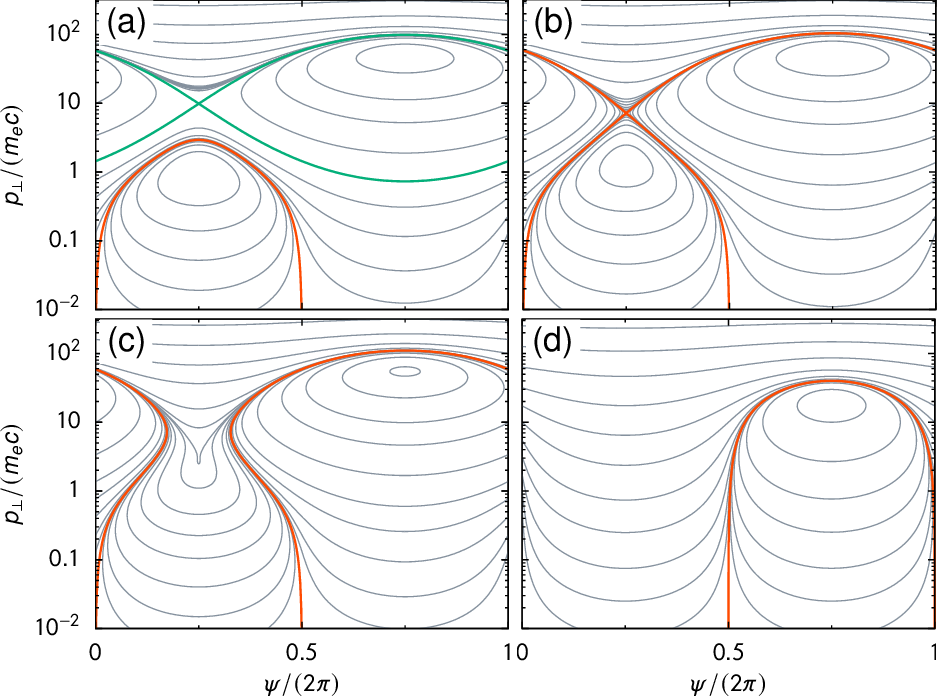}%
\caption{
(a-c) Electron trajectories in the phase-momentum diagram that described as the contours of the Hamiltonian given by Eq.~(\ref{eq:ham}).
The characteristics of the electron motions are determined by two parameters $A = 2 (1+R) a_0$ and $B = 2 B_{\rm ext} / B_c$, which denotes the standing wave amplitude and the external magnetic field strength, respectively.
These parameters are (a) $A = 40$ and $B = 60$, (b) $A = A_1 = 45.043$ and $B = 60$,  and  (c) $A = A_2 = 50.924$ and $B = 60$.
The red and green curves in these figures indicate the separatrixes for the electron trajectory.
(d) Electron trajectories when the external magnetic field is absent, which are given by Eq.~(\ref{eq:ham2}).
The parameters for this case are $A = 40$ and $B = 0$.
\label{fig6}}
\end{figure*}

The contour lines of the Hamiltonian denote the electron trajectories in $\psi$-$\widetilde{p}_{\perp}$ space, some of which are shown in Fig.~\ref{fig6}.
Fig.~\ref{fig6}(a) shows the case of $A = 40$ and $B = 60$ which represents when the wave amplitude is relatively small.
Two fixed points with non-relativistic and relativistic momenta are manifested as the central point of a closed orbit at $\psi = \pi / 2$ and $3 \pi / 2$.
The non-relativistic and relativistic orbits are separated by the separatrixes indicated by the red and green lines in the figure.
Obviously, electrons with a non-relativistic velocity cannot move into the relativistic regime.
This condition is the same as the one discussed by Matsukiyo \& Hada \cite{matsukiyo09}.

As the amplitude of the standing wave increases, the electron orbital motions undergo bifurcations.
Then non-relativistic electrons are permitted to be accelerated to relativistic velocities \cite{isayama23}.
Figure~\ref{fig6}(b) shows the electron trajectories at the condition where the first bifurcation occurs ($A \equiv A_1$).
Now, the boundary of the non-relativistic orbits is connected to the relativistic part through the crossing point at $\psi = \pi / 2$.
Thus, it becomes possible for all the particles at $\pi < \psi < 2 \pi$ to have a relativistic velocity.
The longitudinal momentum cannot be ignored when $\widetilde{p}_{\perp} \gtrsim 1$.
In such a regime, the hot electrons are known to behave like free particles while maintaining their relativistic velocity \cite{isayama23}.
However, for $0 < \psi < \pi$, there still remains non-relativistic closed orbits in this case.

The value of $H$ at the non-relativistic separatrix is calculated as $H (0, \pi) = -B$.
The condition for the bifurcation at $A = A_1$ is equivalent that there is only one $p_{\perp}$ satisfing $H = - B$ at $\psi = \pi / 2$.
The equation of $H ( \chi, \pi / 2 ) = - B$ corresponds to the following cubic equation for $\widetilde{p}_{\perp}$,
\begin{equation}
\widetilde{p}_{\perp}^3 + 2 A \widetilde{p}_{\perp}^2 
+ (A^2 - B^2 +2 B ) \widetilde{p}_{\perp} + 2 AB = 0 \;.
\label{eq:h1}
\end{equation}
Thus, the bifurcation condition can be derived as the discriminant of Eq.~(\ref{eq:h1}) becoming zero.
Eventually, the exact formula for $A_1$ is solved as
\begin{eqnarray}
A_1 &=& \frac{\sqrt{2}}2 
\Bigl[ \left( 2 B^2 + 10 B - 1 \right) \nonumber \\
&& \left. - \left( 64 B^3 + 48 B^2 + 12 B + 1 \right)^{1/2}
\right]^{1/2} \;.
\label{eq:a1}
\end{eqnarray}
For $B=60$, the first bifurcation occurs at $A_1 = 45.043$.

As the intensity of the standing wave increases furthermore, the second bifurcation makes the trajectories of all non-relativistic electrons ($\widetilde{p}_{\perp} \ll 1$) at any phases $\psi$ connected to the relativistic velocity.
The trajectories at the transition ($A \equiv A_2$) are shown in Fig.~\ref{fig6}(c).
Then, if the wave amplitude is $A > A_2$, all electrons in the non-relativistic regime can acquire relativistic energy without the injection problem.
In the PIC simulations, the ponderomotive force concentrates the electrons at the trough of the magnetic field in the preplasma region.
Then finally, entire electrons at any position $x$ have been perfectly accelerated to relativistic velocity within the relativistic gyration timescale $\gamma / \omega_{ce}$ \cite{isayama23}.

The conditions for the second bifurcation can also be derived analytically with the following prescription.
Look at the $H$ value profile along a line of $\psi = \pi / 2$.
In order to ensure no closed orbit at all, $H(\widetilde{p}_{\perp}, \pi / 2)$ must be a monotonically increasing function of $\widetilde{p}_{\perp}$ in the range of $\widetilde{p}_{\perp} > 0$.
Then, the condition can be recognied that $\partial H(\widetilde{p}_{\perp}, \pi / 2) / \partial \widetilde{p}_{\perp}$ must be positive for $\widetilde{p}_{\perp} > 0$.
It can be obtained by setting the discriminant derived from $\partial H(\widetilde{p}_{\perp}, \pi / 2) / \partial \widetilde{p}_{\perp} = 0$ to zero.
By calculating the discriminant of the following fourth-order equation,
\begin{equation}
4 \widetilde{p}_{\perp}^4 + 4 A \widetilde{p}_{\perp}^3
+ (A^2 - B^2 + 4 ) \widetilde{p}_{\perp}^2 
+ 4 A \widetilde{p}_{\perp} + A^2 = 0 \;,
\label{eq:dhdp}
\end{equation}
the critical amplitude is solved simply as 
\begin{equation}
A_2 = 2 \left[ \left( \frac{B}2 \right)^{2/3} - 1 \right]^{3/2} \;, 
\label{eq:a2}
\end{equation}
which corresponds to $A_2 = 50.924$ for $B = 60$.
In the limit of $\widetilde{B}_{\rm ext} \gg 1$, the critical amplitude is given by $A_2 \approx B$, so that the relation $\widetilde{B}_{\rm ext} \lesssim (1 + R) a_0$ will be an excellent indicator of the occurrence of the relativistic two-wave resonant acceleration of electrons.

\begin{figure}
\includegraphics[scale=0.85,clip]{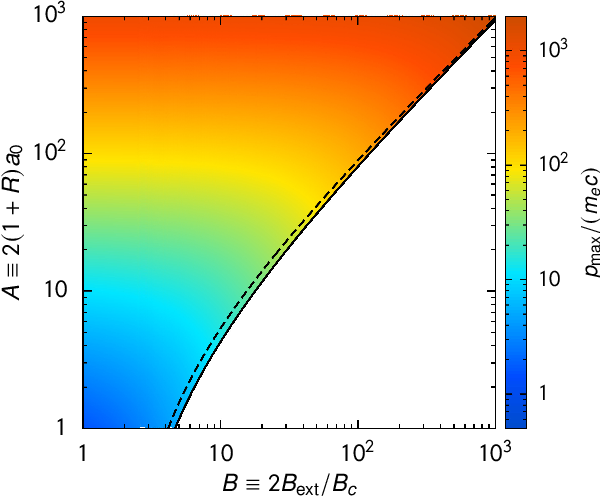}%
\caption{
Theoretical prediction of the maximum electron momentum due to the relativistic two-wave resonant acceleration in counter-propagating circularly polarized waves.
The color map shows the maximum momentum $p_{\max}$ obtained as the root of Eq.~(\ref{eq:pmax}) depending on two parameters $A = 2 (1+R) a_0$ and $B = 2 B_{\rm ext} / B_c$.
The black solid and dashed lines indicate the boundaries for the bifurcations in the electron trajectory $A = A_1$ and $A = A_2$, respectively.
\label{fig7}}
\end{figure}

The peak value of $\widetilde{p}_{\perp}$ along the separatrix is taken at $\psi = 3 \pi / 2$.
Then, the maximum momentum $\chi_{\max} = \widetilde{p}_{\max}^2$ is obtained solving the equation of $H (\chi_{\max}, 3 \pi / 2) = - B$, that is,
\begin{equation}
\widetilde{p}_{\max}^3 - 2 A \widetilde{p}_{\max}^2 
+ (A^2 - B^2 +2 B ) \widetilde{p}_{\max} - 2 AB = 0 \;.
\label{eq:pmax}
\end{equation}
Figure~\ref{fig7} depicts the maximum momentum evaluated as a root of Eq.~(\ref{eq:pmax}) in the region where the transition to relativistic energy is possible.
The bifurcation boundaries $A = A_1$ and $A = A_2$ are drawn with the solid and dashed lines.
The maximum momentum depends mainly on the amplitude of the standing wave.
For a given $A$, the larger the magnetic field strength $B$ is, the larger the maximum energy of electron becomes.
Therefore, for maximizing the relativistic two-wave resonant acceleration,
 the parameters $A$ and $B$ should be closer to the bifurcation boundary $A \sim B$, that is, $a_0 \sim \widetilde{B}_{\rm ext}$.
This fact is consistent with the electron energy peak in our PIC simulations shown in Fig.~\ref{fig4}(c).
The total energy is proportional to the injected laser energy $a_0^2$ since the conversion rate when $a_0 = \widetilde{B}_{\rm ext}$ takes a similar value (see Table~\ref{tab1}).

The approximate formula for $\widetilde{p}_{\max}$ is expressed as
\begin{equation}
\widetilde{p}_{\max} \approx A + \left[ B ( B - 2 ) \right]^{1/2} \;,
\end{equation}
in the limit of $\widetilde{p}_{\max} \gg \sqrt{B}$. 
This simple formula would be useful for quick estimation because the deviation from the exact solution is less than 3\% in the region of $B \gtrsim 10$.
The maximum momentum given by Eq.~(\ref{eq:pmax}) is, in fact, valid not only when $\widetilde{B}_{\rm ext} > 1$, but also when the magnetic field is weaker.
When $\widetilde{B}_{\rm ext} = 0$, the Hamiltonian is modified as
\begin{equation}
H (\chi, \psi) = A \chi^{1/2} \sin \psi + \chi \;.
\label{eq:ham2}
\end{equation}
The trajectory analysis confirms that the maximum momentum of this unmagnetized case can be written as
\begin{equation}
\widetilde{p}_{\max} = A \;.
\end{equation}

The analytical solutions of the maximum momentum could be related to the maximum electron energy evaluated by our numerical simulations.
The normalized kinetic energy of electrons, $\widetilde{\epsilon} =
{\epsilon}/m_e c^2$, corresponding to the maximum momentum is given by
\begin{equation}
\widetilde{\epsilon}_{\max} = 
(1 + \widetilde{p}_{\max}^2)^{1/2} - 1 \;, 
\label{eq:emax}
\end{equation}
which is shown in Fig.~\ref{fig4}(a) assuming the reflectivity $R = (N - 1)/(N + 1)$.
The edge of the bifurcation at $\widetilde{B}_{\rm ext} = 56.9$ given by Eq.~(\ref{eq:a1}) is perfectly consistent with a sharp drop in the electron energy, which assures the accuracy of the theoretical picture.
The predicted energy gives the same order of the simulation results for $\epsilon_{e,\max}$, but is always below them at most a factor of a few.
The difference would be because the theoretical prediction is evaluated under the assumption of $\widetilde{p}_{\parallel} = 0$, while the actual acceleration eventually increases the longitudinal momentum as well.

The temperature of bulk electrons is much lower than the hot electron energy by many orders. 
Here, we attempt to estimate the bulk temperature by a simplified consideration.
The resistive heating is expected to be the dominant process for electron heating in our situation.
Then the energy equation is given by
\begin{equation}
\frac32 \frac{\partial}{\partial t} (k_B T_e) \approx 
m_e \nu_{ei} | \bm{v}_e - \bm{v}_i |^2 \;,
\end{equation}
where $k_B$ is the Boltzmann constant.

If $\widetilde{B}_{\rm ext}$ is greater than unity, the relative velocity between electrons and ions is determined by the quiver motion of the whistler wave \cite{sano19}.
Assuming the Spitzer formula for the collision frequency \cite{chen84},
\begin{equation}
\nu_{ei} = \frac{\ln \Lambda}{3 (2 \pi)^{3/2}}
\frac{Z e^4}{\epsilon_0^2 m_e^{1/2}}
\frac{n_e}{(k_B T_e )^{3/2}} \;,
\end{equation}  
the bulk electron temperature 
scaling with $\widetilde{n}_e$, 
$\widetilde{B}_{\rm ext}$, $a_0$, and time $\widetilde{t}$ is derived as 
\begin{equation}
\frac{k_B T_e}{m_e c^2} \sim
\left[
\frac{40 (2 \pi)^{3/2} \ln \Lambda}{9} 
\frac{Z r_e}{\lambda_0}
\frac{\widetilde{n}_{e} a_0^2}
{(N+1)^2 (\widetilde{B}_{\rm ext} -1)^2}
 \widetilde{t}
\right]^{2/5} 
\label{eq:tbulk}
\end{equation}  
\cite{sano19}, where $\ln \Lambda$ is the Coulomb logarithm, $Z$ is the ion charge number, and $r_{e} = e^2 / (4 \pi \epsilon_0 m_e c^2)$ is the electron classical radius.
This approximation is valid when the quiver velocity is non-relativistic, $v_q \lesssim c$, so that the relation
\begin{equation}
\frac{2 a_0}{(N+1)(\widetilde{B}_{\rm ext} - 1)} \lesssim 1
\end{equation}
should be satisfied.

When the external magnetic field is below the critical $B_c$, the whistler wave cannot enter the target. 
Then the bulk electrons are heated by the collisions with the return current. 
Using an assumption $\gamma n_c c = n_e v_{\rm ret}$ and the ponderomotive scaling for hot electron energy $\gamma = ( 1 + a_0^2 )^{1/2}$, the return current velocity $v_{\rm ret}$ is estimated as
\begin{equation}
\frac{v_{\rm ret}}{c} \sim \frac{( 1 + a_0^2 )^{1/2}}{\widetilde{n}_e} \;.
\end{equation}
Then the bulk temperature due to the return current is obtained by
\begin{equation}
\frac{k_B T_e}{m_e c^2} \sim
\left[
\frac{20 (2 \pi)^{3/2} \ln \Lambda}{3} 
\frac{Z r_e}{\lambda_0}
\frac{(1 + a_0^2)}
{\widetilde{n}_e}
 \widetilde{t}
\right]^{2/5} \;.
\label{eq:tbulk2}
\end{equation}

The bulk electron temperatures predicted by Eqs.~(\ref{eq:tbulk}) and (\ref{eq:tbulk2}) are overplotted in Fig. \ref{fig4}(a).
When whistler waves cannot penetrate the target, the bulk temperature is independent of the external magnetic field, as seen from Eq.~(\ref{eq:tbulk}).
On the other hand, for $\widetilde{B}_{\rm ext} \gtrsim 1$, the electrons in the target is heated due to the electron quiver motion of whistler wave [Eq.~(\ref{eq:tbulk2})].
In the range where the non-relativistic approximation holds ($\widetilde{B}_{\rm ext} \gtrsim 32$), the PIC results are in excellent agreement with the theoretical model.
In the region where the relativistic amplitude must be taken into account ($1 \lesssim \widetilde{B}_{\rm ext} \lesssim 32$; dashed thick line), there is a slight discrepancy from the theory, but the accuracy would be still adequate for the use of the order estimation.

\subsection{The optimal condition for hot electron fraction}

\begin{figure*}
\includegraphics[scale=0.85,clip]{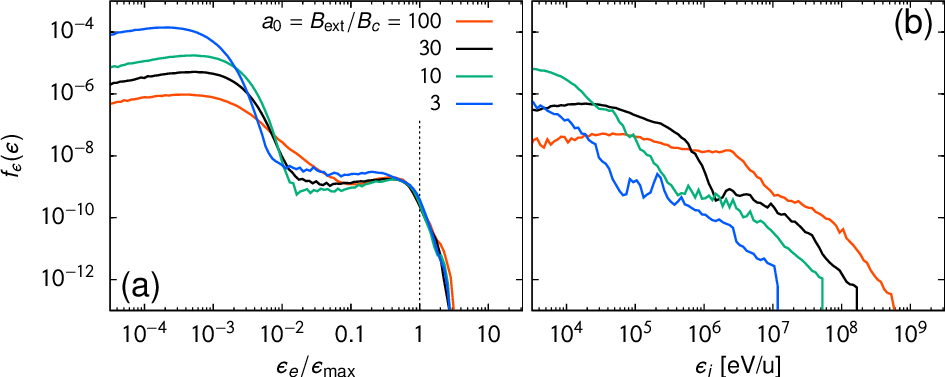}%
\caption{
(a) Electron energy spectra and (b) ion energy spectra obtained under the optimal condition of $a_0 = B_{\rm ext} / B_c$.
The spectra correspond to the cases of $a_0 = 100$ (red; run10), 30 (black; run01), 10 (green; run09), and 3 (blue; run08).
The horizontal axis is normalized by the theorerical maximum energy for each cases given by Eqs.~(\ref{eq:pmax}) and (\ref{eq:emax}).
Notice that the unit of the energy is eV/u for the ion spectra.
\label{fig8}}
\end{figure*}

The condition under which the relativistic two-wave resonant acceleration of electrons is most efficient has derived in the previous subsection as $A \sim B$, or $a_0 \sim \widetilde{B}_{\rm ext}$.
We will now check the energy spectra of charged particles under this optimized condition.
The energy distributions of electrons for various laser intensities are shown in Fig.~\ref{fig8}(a).
Here the electron kinetic energy is normalized by the maximum energy $\epsilon_{\max}$ evaluated from Eqs.~(\ref{eq:pmax}) and (\ref{eq:emax}).
In all cases, a considerable amount of hot electrons are generated through the electron acceleration at the standing wave, resulting in the two components in the energy spectra.
The peak of hot electrons is comparable to the energy for the maximum momentum.
In the limit of $a_0 = \widetilde{B}_{\rm ext}\gg 1$, it corresponds to $\widetilde{p}_{\max}\sim 4 \widetilde{B}_{\rm ext}$. 
Thus, the peak energy increases approximately in proportion to $\widetilde{B}_{\rm ext}$.
On the other hand, the temperature of the bulk electrons is given by Eq.~(\ref{eq:tbulk}), which should be proportional to $\widetilde{B}_{\rm ext}^{2/5}$.
The electron spectra shown in Fig.~\ref{fig8}(a) strongly support the validity of the analytical estimations on the parameter dependence.

\begin{figure*}
\includegraphics[scale=0.85,clip]{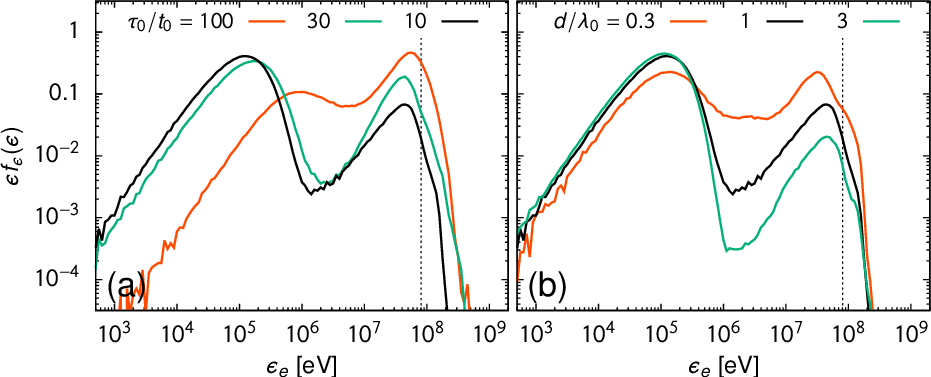}%
\caption{
(a) Effects of the laser pulse duration on electron energy spectra demonstrated by three runs of $\tau_0 / t_0 = 100$ (red; run15), 30 (green; run14), and 10(black; run01).
(b) Comparison of three runs with different target thicknesses, $d / \lambda_0 = 0.3$ (red; run18), 1 (black, run01), and 3 (green; run19).
The fiducial parameters are used in all these one-dimensional calculations  ($a_0 = 30$ and $B_{\rm ext} / B_c = 30$).
The dotted lines in both panels correspond to the theoretical prediction of the maximum energy.
\label{fig9}}
\end{figure*}

For the electron acceleration using standing waves, the number of hot electrons must be enhanced if the duration of the standing wave becomes longer.
Figure~\ref{fig9}(a) indicates the relationship between the pulse duration and the hot electron features.
These spectra are obtained when the laser pulse is extended by a factor of 3 and 10 from the fiducial case.
To account for the longer pulse length, the measured timing of the spectrum is reasonably late, and the computational domain is enlarged by the same factor.
As expected, the fraction of hot electrons increases with increasing the pulse length.
The longer the duration of the standing wave, the more chance the electrons will move to the target surface and be accelerated.
For the case of $\widetilde{\tau}_0 = 100$, the hot electrons overwhelm the bulk component by number.
The ratio of the hot electron number to the total increases monotonically, which is 0.11, 0.25, and 0.73 for $\widetilde{\tau}_0 = 10$, 30, and 100, respectively, as listed in Table~\ref{tab1}.
Further extension of the duration would enable converting all the electrons in the target to relativistic energy.
Thus, a longer pulse duration of the incident laser would be a great advantage for this acceleration mechanism.

From the point of efficiency, it would make sense to have an appropriate target thickness that matches the duration of the standing wave.
Figure~\ref{fig9}(b) shows the electron energy distributions of varying the thickness of the target.
The pulse length is fixed as $\widetilde{\tau}_0 = 10$ for all the runs.
Thus, there is almost no difference in the generated number of hot electrons.
Instead, looking at the fraction of hot electrons, it decreases with increasing thickness, that is 0.457, 0.106, and 0.029 for $\widetilde{d} = 0.3$, 1, and 3 (see Table~\ref{tab1}).
It can be understood that the difference is originated from the increase of unaccelerated electrons inside the thicker target.
Therefore, the relation between the target thickness and laser pulse length is genuinely essential for optimization.

\section{Discussion \label{sec4}}

\subsection{Ion acceleration by enhanced sheath fields}

According to innovative progress in available laser intensity, laser-driven ion acceleration has become an attractive technique for the purposes such as plasma diagnostics and medical applications \cite{daido12}. 
Various mechanisms of ion acceleration have been proposed theoretically and attempted in laboratories.
Among them, the most straightforward and robust mechanism is called the target normal sheath acceleration (TNSA) \cite{wilks01}.
By irradiating a thin foil with an intense laser beam, the electric sheath field is formed naturally by the ejection of relativistic electrons from the backside of the target.
Ions are pulled out of the target by the sheath field and accelerated due to the electric potential gap.

Since the size of the sheath field is affected severely by the temperature (or the average energy) of hot electrons, the electron energy distribution is crucial for the maximum energy of the accelerated ions \cite{mora79,mora03}.
It is known that the number density of hot electrons also contributes to the longer acceleration time \cite{fuchs07}.
Furthermore, a model fitting of the experimental achievement supports the idea that both the temperature and density of hot electrons are essential variables in determining the maximum ion energy \cite{takagi21}.
Thus, the relativistic two-wave resonant acceleration in standing whistler waves may dramatically alter the ion acceleration by TNSA. 
In this subsection, we will briefly quantify how the characteristics of the accelerated ions are modified by the external magnetic field and the laser intensity.

Figure~\ref{fig8}(b) shows the energy spectrum of carbon ions obtained in the optimized runs with $a_0 = \widetilde{B}_{\rm ext}$.
The maximum energy increases roughly in proportion to the laser amplitude, for instance, $\epsilon_{i,\max} = 11$ MeV/u for $a_0 = 3$ and $5.7 \times 10^2$ MeV/u for $a_0 = 100$.
The energy conversion rate from the incident laser to ions is always 20--30\% in all cases.
Whereas the average energy is enhanced by many orders from $\langle \epsilon_i \rangle = 1.0 \times 10^5$ eV/u for $a_0 = 3$ to $\langle \epsilon_i \rangle = 1.4 \times 10^8$ eV/u for $a_0 = 100$.
Ion acceleration utilized by our mechanism could be a promising option for future production schemes because the required carbon-ion energy for medical applications is about 400 MeV/u.

By the way, if $\widetilde{B}_{\rm ext}\gtrsim a_0$, the resonant acceleration of electrons has no chance, but ion heating due to the excitation of ion compressible waves becomes efficient instead \cite{sano19,sano20a,hata21}. 
Therefore, such a regime would be suitable for generating fusion plasmas by direct energy transfer from the electromagnetic waves to ions.

\begin{figure}
\includegraphics[scale=0.85,clip]{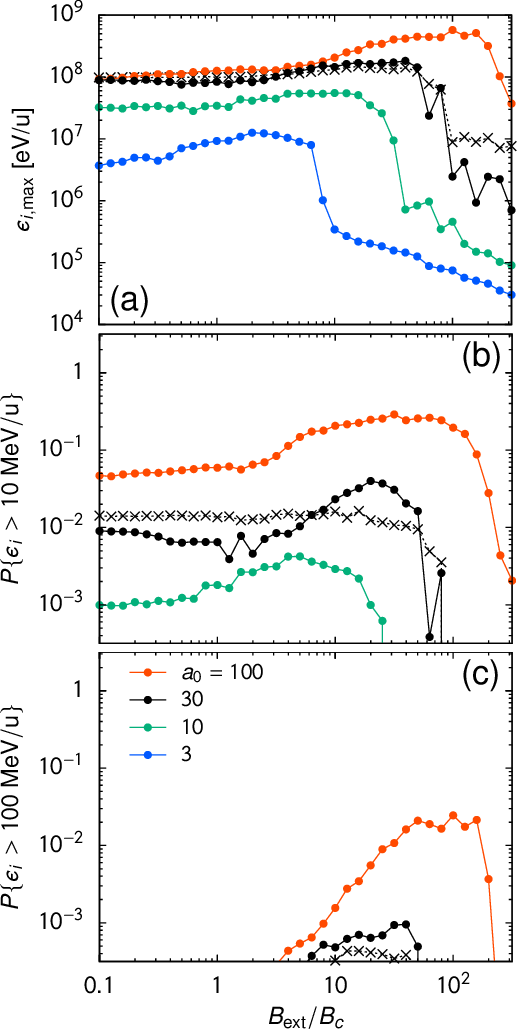}%
\caption{
(a) The maximum ion energy and the number fraction of carbon ions over a threshold energy of (b) $\epsilon_{\rm thr} = 10$ MeV/u, and (c) 100 MeV/u in a series of runs with various $B_{\rm ext} / B_c$ and $a_0$.
The meanings of the marks are identical to those of Fig.~\ref{fig4}(b).
\label{fig10}}
\end{figure}

The dependence of the maximum ion energy on $\widetilde{B}_{\rm ext}$ resembles that of electrons [see Figs.~\ref{fig10}(a) and \ref{fig4}(a)].
For the range of $1 \lesssim \widetilde{B}_{\rm ext} \lesssim a_0$, where the acceleration condition is satisfied, the maximum energy seems to be higher than the case without the magnetic field, but the difference is less than an order of magnitude.
There is no quantitative difference with the case of LP laser irradiation.
Thus, the maximum ion energy does not benefit from this electron acceleration mechanism by much.
On the other hand, a significant advantage exists in the number fraction of accelerated ions.
Figures~\ref{fig10}(b) and \ref{fig10}(c) show the ion fraction above 10 MeV/u and 100 MeV/u, respectively.
The magnetic field dependence inherits the characteristics of the electron acceleration.
For $a_0 = 100$, a large fraction of ions more than a few 10\% are achieved for $\epsilon_{i,{\rm thr}} = 10$ MeV/u in a wide range of the field strength, $3 \lesssim \widetilde{B}_{\rm ext}\lesssim 100$.
Furthermore, the ion fraction over 100 MeV/u becomes a few percent of total ions if the conditions are right (see also Table~\ref{tab3}).

In the standing waves of LCP waves, the ion version of the two-wave resonant acceleration can be expected in principle \cite{isayama23}.
However, the required magnetic field strength for the ion cyclotron resonance is much larger than the electron case by a factor of the mass ratio $m_i / m_e$.
Thus, for the laboratory application, such field strength is far beyond the available range at present.

\subsection{Influence of Preplasma Density Profile}

In this work, we focus on the electron acceleration in the preplasma region.
The presence of preplasma cannot be ignored when considering energy conversion processes near the target surface, especially in the cases of intense lasers.
However, there is a quantitative uncertainty in preplasma properties (e.g., density and scale length) in actual experiments.
In our simulations, we fixed the scale length as $\lambda_0$, and the maximum density of the preplasma at the target surface as $10 n_c$.
The dependence of the plasma density on the acceleration mechanism by standing whistler waves is not so significant as can be understood by the test particle analysis.
We have confirmed that the same acceleration occurs even with the reduced density by an order of magnitude or different scale lengths.
Therefore, it can be said that the effect of the density distribution of the preplasma is unimportant and that electron acceleration by standing whistler waves is a universally occurring phenomenon in underdense preplasma.

\subsection{Laser Incident Angle}

So far, we have considered only the setup where the magnetic field and laser propagation directions are perfectly coincident.
However, other situations in which those directions are misaligned will occur in both space phenomena and laboratory experiments.
Then, whether the laser-plasma interaction is affected by shifting the magnetic field angle is an interesting question.

Suppose that the incident angle of the laser is fixed so that it is perpendicular to the target surface.
The angle of the magnetic field to the laser direction is denoted by $\theta$.
The one-dimensional simulations with various $\theta$ reveal that the energy distributions of electrons and ions are almost unchanged if the angle is less than 15 degrees.
Therefore, a slight angle deviation would not be a serious problem when trying to verify this acceleration mechanism experimentally.
If the angle becomes around $\theta = 45$ deg, the energy conversion efficiency to plasma is reduced by about 20\%, which may be a level that cannot be neglected (see run22--run26 in Table~\ref{tab1}).
Other wave modes such as the X-wave must be considered when the magnetic field has a perpendicular component to the laser propagation.
Such a complex interaction would be worth analyzing in detail in the future.

\subsection{Multi-dimensional effects}

Once a standing wave forms, the electron acceleration process in this study would be independent of the spatial dimension.
We have verified it by running two-dimensional simulations.
The computational box size of the additional spatial direction $y$ is set to $\widetilde{L}_y = 30$, which is sufficiently broader than the laser focal spot $3 \lambda_0$ of the Gaussian shape.
The target and preplasma conditions are the same as in the one-dimensional fiducial run.
The key parameters are taken as $a_0 = \widetilde{B}_{\rm ext} = 30$, and the polarization of the laser light is right-hand circular.
In order to demonstrate multi-dimensional effects, the incident angle of the laser irradiation to the $x$ axis is assumed to be $\varphi = 26.6$ deg ($\tan \varphi = 0.5$), whereas the external magnetic field is along the $x$ direction.
In the $y$ direction, we adopt the periodic boundary condition for simplicity.
The resolution in two-dimensional runs is $\varDelta x = c \varDelta t = \lambda_0 / 200$, and the initial particle number per cell is 20 and 120 for ions and electrons, respectively.

\begin{figure}
\includegraphics[scale=0.85,clip]{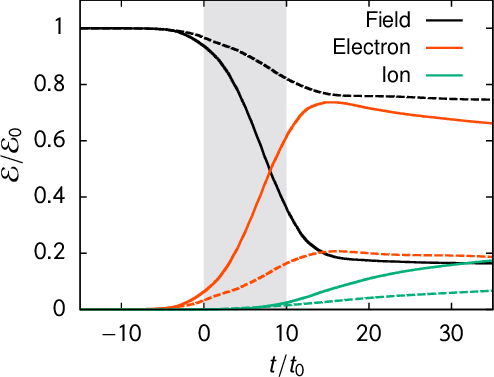}%
\caption{
Time histories of the field and plasma energies in the two-dimensional run.
The model parameters are identical to those in the one-dimensional fiducial run ($n_e / n_c = 603$, $B_{\rm ext} / B_c = a_0 = 30$).
A right-hand circularly polarized laser is injected toward a carbon foil with an incident angle $\varphi = 26.6$ deg.
The color indicates the energy of the electrons (red), ions (green), and electromagnetic waves (black). 
For comparison, the results of the case without the external magnetic field are also plotted by the dashed curves. 
The gray area denotes the laser pulse duration, where the injected laser irradiates the target surface ($0 \le t/t_0 \le 10$). 
\label{fig11}}
\end{figure}

\begin{figure*}
\includegraphics[scale=0.85,clip]{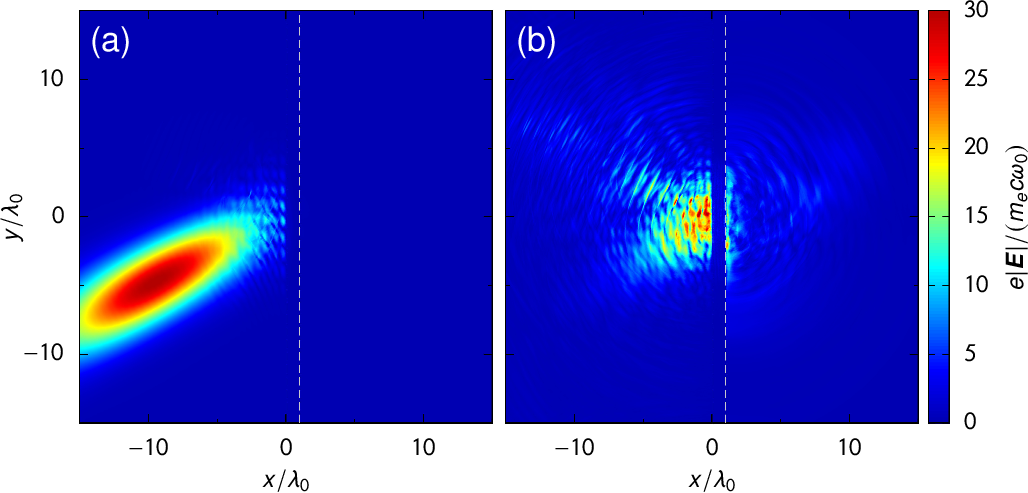}%
\caption{
Spatial distributions of the electric field $| \bm{E} | / E_c$ where $E_c = m_e c \omega_0 / e$.
The snapshot data are taken at (a) $t / t_0 = -4$ and (b) $t / t_0 = 11$.
The right-hand circularly polarized laser comes from the left with an incident angle $\varphi = 26.6$ deg. 
The carbon target surface is $x = 0$ initially, and the thickness is $d / \lambda_0 = 1$, indicated by the gray dashed lines.
The feature of standing waves can be recognized by periodic stripes parallel to the target surface in this image.
\label{fig12}}
\end{figure*}

Figure~\ref{fig11} shows the time history of the energy conversion.
The injected laser energy is transferred to electrons and reduced by about 80\% immediately in the timescale of pulse duration $\widetilde{\tau}_0 = 10$.
Although the laser light has a finite incident angle $\varphi$, a standing wave appears near the target surface (see Fig.~\ref{fig12}).
Electrons are accelerated effectively by two-wave resonance in the standing wave, as seen in the one-dimensional cases.
The ion energy increases gradually after the electron acceleration and sheath formation.
Then, the electrons acquire almost all the energy of the electromagnetic wave during the direct interaction.
It can be seen in Fig.~\ref{fig12}(b) where the injected electric field is extinguished while forming a standing wave at the end of the laser pulse 
$\widetilde{t} \sim \widetilde{\tau}_0$.

\begin{figure*}
\includegraphics[scale=0.85,clip]{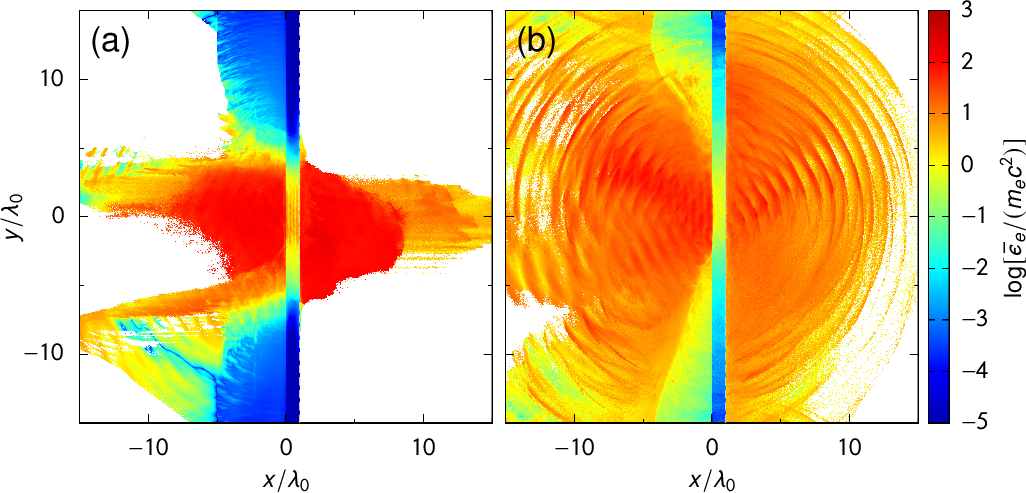}%
\caption{
Spatial distributions of the electron average energy $\bar{\epsilon}_e / (m_e c^2)$ for the cases (a) with an external magnetic field ($B_{\rm ext} / B_c = 30$) and (b) with no field ($B_{\rm ext} = 0$).
The orientation of the magnetic field is parallel to the $x$ axis or the target-normal direction.
The other model parameters are the same as in Fig.~\ref{fig12}.
The time taken snapshot data is $t / t_0 = 11$ for both cases.
\label{fig13}}
\end{figure*}

The spatial distributions of the accelerated electrons are displayed by the average kinetic energy $\bar{\epsilon}_e$ in each cell.
The relativistic electrons are trapped by the Larmor motions in the $y$ direction when the external magnetic field exists [Fig.~\ref{fig13}(a)].
The Larmor radius for this case is $\widetilde{r}_L = ({2 \pi \widetilde{B}_{\rm ext}})^{-1}$,
which is much shorter than the laser wavelength.
Thus, there is no structure in the tangential direction to the target surface, which preserves the phenomenon's one-dimensionality.
The amplitude of the sheath field is also similar to the one-dimensional cases.
A strong magnetic field provides a significant advantage by confining accelerated electrons, as without a magnetic field, they would diverge isotropically [Fig.~\ref{fig13}(b)].
The accelerated electrons spread out in the $x$ direction much wider than the original target thickness of $\widetilde{d} = 1$.
Those electrons are recirculated between the front and rear surfaces.
A strong sheath field forms at both sides of the target, and this electric field accelerates ions.

\begin{figure}
\includegraphics[scale=0.85,clip]{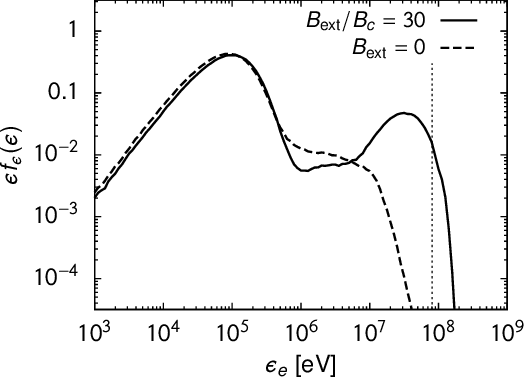}%
\caption{
Energy spectra of electrons obtained from PIC simulations taking account of two-dimensional geometry.
Except for the dimensionality, the model parameters and plot marks are the same as Fig.~\ref{fig1}(a).
The spectra are calculated from the electrons in a $|y / \lambda_0 |
\le 1.5$ region at the end of calculation $t / t_0 = 34$.
The dotted lines correspond to the theoretical prediction of the maximum energy given by Eqs.~(\ref{eq:pmax}) and (\ref{eq:emax}).
\label{fig14}}
\end{figure}

The electron energy distributions plotted in Fig. \ref{fig1}(a) are unaltered even in two dimensions.
Figure~\ref{fig14} shows the electron spectra taking account of the spatial multi-dimensionality.
Notice that the electrons only within $| \widetilde{y} | \le 1.5$ are considered for the spectrum.
The characteristics of hot electrons around 10--100 MeV are hardly changed.
The bulk temperature is slightly higher due to the lower resolution than the one-dimensional cases, but it does not affect the hot electron features.
Therefore, it can be verified that two-wave resonance acceleration in standing whistler waves also occurs in multi-dimensional geometries, suggesting that it is a universal and robust acceleration mechanism.

In our two-dimensional simulations, standing waves could exist adequately in space and time as the acceleration site.
In the absence of a magnetic field, the interaction of a relativistic laser with an overdense target has been reported to generate surface modulation, such as filamentary structures in the density and self-generated magnetic field \cite{kemp09,may11,huller19}.
The interface fluctuation will have a non-negligible influence on forming standing waves consisting of incident and reflected waves.
Indeed, surface structure modulation within a scale of the laser spot
size has been observed in our simulations without an external magnetic field.
In that case, filamentary density stripes are formed in the direction of laser propagation.
However, the interface remains relatively flat, and no significant modulation is seen when applying a strong magnetic field.
This significant difference may be caused by the advent of the whistler mode, which modifies the transmittance properties of the intense laser light.
A comprehensive understanding of complex surface modulation with and without an external magnetic field will need a more thorough investigation in the future.

\section{Conclusions \label{sec5}}

We have investigated laser-plasma interaction in the presence of a strong magnetic field and revealed the physical mechanism and optimal conditions for hot electron generation.
Electrons are accelerated efficiently in standing waves formed by two RCP electromagnetic waves that counter-propagate along an external magnetic field. 
The standing wave of CP lights and the external magnetic field are essential ingredients for this acceleration mechanism.
If these two elements coexist, the energy conversion process from electromagnetic waves to electrons through wave-particle interaction will result in a completely different picture than the conventional one.
Our findings are summarized as follows.

(i) The structure of the electromagnetic fields in the CP standing wave is intrinsic to this electron acceleration.
The cyclotron resonance caused by the background magnetic field quickly raises the electron velocity from non-relativistic to relativistic.
The magnetic field must be strong such that the electron cyclotron frequency is larger than the laser one.

(ii) Furthermore, for this acceleration to be realized, there is the minimum requirement in the amplitude of the electromagnetic wave.
The condition for the bifurcation in the electron gyration motion is that the magnetic field amplitude of the standing wave must be larger than the external magnetic field.

(iii) The stronger the external magnetic field, the higher the maximum energy of the accelerated electrons. Therefore, there is the optimal condition for the electron acceleration, which is when the external magnetic field $\widetilde{B}_{\rm ext}$ is roughly equal to the injected laser amplitude $a_0$.

(iv) Generation of a  large number of hot electrons brings a significant advantage in laser-driven ion acceleration.
For the case of carbon ions, the maximum energy reaches a few hundreds of MeV/u, and the number fraction over 10 MeV/u could be more than 10 percent. 

(v) This relativistic two-wave resonant acceleration occurs even in two-dimensional geometry as well as the ideal one-dimensional situation. 
A slight deviation in the angle between the laser and magnetic field has little influence on the electron acceleration, indicating the robustness of this acceleration process.

The experimental verification of this mechanism will be a precious step in the future.
The laser conditions used in this study can be achievable with a realistic TW-class femtosecond laser.
The most challenging part is to prepare a uniform magnetic field much stronger than $B_c \approx 10$ kT.
Recently, laser-driven magnetic fields have been generated in various ways \cite{yoneda12,fujioka13,korneev15,goyon17,santos18,morita23,shi23}, but they are yet in the order of kT.
Theoretically, the generation of even stronger magnetic fields of MT-class has been proposed by designing nm-scale structural targets \cite{murakami20,shokov21,zosa22}.
Since the synchronization of the field orientation with the laser direction is also essential, the self-generation of the axial magnetic field has to be well controlled, which may raise the difficulty even more.
However, strong magnetic fields bring drastic and attractive differences in laser-plasma interaction.
Thus, it would be worth challenging to pursue such extreme magnetic fields because there are many practical applications of hot electron generation, such as laser-driven ion acceleration and fusion plasma generation.

\begin{acknowledgments}
We thank Y. Sakawa, Y. Sentoku, and K. Sugimoto for valuable discussions.
This work was performed under the joint research project of the
Institute of Laser Engineering, Osaka University.  
This work was partly achieved through the use of SQUID at the Cybermedia Center, Osaka University.
This research was supported by JSPS KAKENHI Grant No. JP24H00204, 
No. JP21K03500, No. JP20H00140, and JSPS Core-to-Core Program,
B. Asia-Africa Science Platforms No. JPJSCCB20190003.
\end{acknowledgments}

\appendix

\section{Parameters and resulted quantities of performed runs \label{app3}}

A list of the simulation results used for the discussions in this paper is summarized in Tables~\ref{tab1}, \ref{tab2}, and \ref{tab3}.
The first column of these tables gives the common label for each run.

Table \ref{tab1} contains the physical parameters for each run. 
The injected laser is characterized by the polarization, the normalized amplitude $a_0$, and the pulse length $\widetilde{\tau}_0$. 
The strength of the external magnetic field is given by $\widetilde{B}_{\rm ext}$, and the angle to the laser direction is $\theta$. 
The target material is carbon plasma in all the runs, and the density is set to $\widetilde{n}_e = 603$. 
The target type is described by the thickness relative to the laser wavelength $\widetilde{d}$ and the presence or absence of preplasma. The energy conversion efficiency to plasmas obtained from each simulation is evaluated regarding the ratio to the incident laser energy. 
The hot electron fraction in the last column is calculated using a procedure explained in Appendix~\ref{app1}.

Table \ref{tab2} summarizes the features of electrons after the laser-plasma interaction.
The maximum and average energies, $\epsilon_{\max}$ and $\langle \epsilon_e \rangle$, are shown in the unit of eV.
This table has three kinds of average energy for the total, bulk, and hot electrons.
The last three columns are the number fraction of electrons exceeding the threshold energy of $\epsilon_{\rm thr} = 1$, 10, and 100 MeV.

Table \ref{tab3} is a list of carbon ion characteristics.
The maximum and average energies are given in the electron volts per nucleon.
The number fractions of ions above the threshold energy of $\epsilon_{\rm thr} = 1$, 10, and 100 MeV/u are listed in the last three columns.

\section{Evaluation of bulk electron temperature \label{app1}}

\begin{figure*}
\includegraphics[scale=0.85,clip]{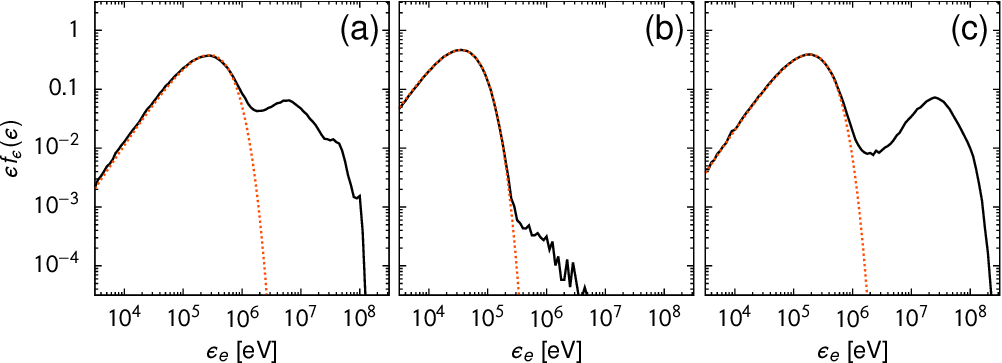}%
\caption{
Fitted results of the bulk electron temperature for the cases of (a) $B_{\rm ext} / B_c = 10$ and $\theta = 0$, (b) $B_{\rm ext} / B_c = 100$ and $\theta = 0$, and (c) $B_{\rm ext} / B_c = 30$ and $\theta = 15$.
The other parameters are identical to the fiducial run.
The bulk component in the energy spectrum is fitted by a thermal distribution of $\alpha f_{\rm MB} (\epsilon, T_{\rm bulk})$.
Here, $f_{\rm MB} (\epsilon, T)$ is the Maxwell-Boltzmann distribution, and $\alpha$ stands for a fraction of bulk electrons.
The best solutions for each spectrum are (a) $T_{\rm bulk} = 184$ keV and $\alpha = 0.83$, (b) $T_{\rm bulk} = 23.4$ keV and $\alpha = 1.0$, and (c) $T_{\rm bulk} = 124$ keV and $\alpha = 0.85$, which are shown by the red dotted curve in each panel.
\label{fig15}}
\end{figure*}

In this analysis, the bulk electron temperature and the average energy of hot electrons are defined based on the electron energy spectrum. 
Here, we describe the procedure and some examples.

The Maxwell-Boltzmann distribution for a non-relativistic temperature $T$ is given by  
\begin{equation}
f_{\rm MB} ( \epsilon, T ) = 2 \pi \left( \frac1{\pi k_B T} \right)^{3/2}
\epsilon^{1/2} \exp \left( - \frac{\epsilon}{k_B T} \right) \;,
\end{equation}
where $\epsilon$ is the kinetic energy of the particles.

The bulk temperature $T_{\rm bulk}$ is obtained by fitting the energy spectrum with a function of $\alpha f_{\rm MB}$, where $\alpha$ ($\le 1$) is a constant factor meaning the fraction of bulk electrons. 
The fitted examples are shown in Fig.~\ref{fig15}. 
The black solid curve depicts the electron spectrum obtained from the PIC calculation $\epsilon f_\epsilon (\epsilon)$, and the dotted red line is the fitted thermal bulk component $\alpha \epsilon f_{\rm MB} (\epsilon, T)$ with $T = T_{\rm bulk}$. 
In the three examples shown here, the characteristics of the hot electrons are all different, but the thermal component is nicely fitted.
By using the bulk temperature, the average energy of the bulk electrons is given by $\langle \epsilon_{e,{\rm bulk}}\rangle = (3/2) k_B T_{\rm bulk}$.

The average energy of hot electrons is estimated from the subtraction of the bulk component from the numerically obtained spectrum.
We integrate the hot electron energy in the range of $\epsilon > k_B T_{\rm bulk}$.
This procedure determines the number fraction and average energy of hot electrons listed in Tables~\ref{tab1} and \ref{tab2}.

\begin{table*}
\caption{Simulated model parameters and the obtained energy conversion rates.
\label{tab1}}
\begin{tabular}{@{\extracolsep{10pt}}lccccccccccc}
\hline \hline
Label & Dim. & Polarity & $a_0$ & $\widetilde{\tau}_0$ & $\widetilde{B}_{\rm ext}$ &
$\theta$ & $\widetilde{d}$ & Preplasma & \multicolumn{2}{c}{Conversion
Rate} & Hot Electron \\ \cline{10-11} 
 & & & & & & [deg] & & & Electron & Ion & Fraction \\ \hline
run01	&	1D	&	RCP	&	30	&	10	&	30	&	0	&	1	&	Yes	&	0.492	&	0.293	&	0.106	\\
run02	&	1D	&	LCP	&	30	&	10	&	30	&	0	&	1	&	Yes	&	0.001	&	0.032	&	0.000	\\
run03	&	1D	&	Linear	&	42	&	10	&	30	&	0	&	1	&	Yes	&	0.186	&	0.143	&	0.033	\\
run04	&	1D	&	RCP	&	30	&	10	&	30	&	0	&	1	&	--	&	0.392	&	0.215	&	0.088	\\
run05	&	1D	&	RCP	&	30	&	10	&	0	&	0	&	1	&	Yes	&	0.009	&	0.081	&	0.007	\\
run06	&	1D	&	Linear	&	42	&	10	&	0	&	0	&	1	&	Yes	&	0.020	&	0.131	&	0.045	\\
run07	&	1D	&	RCP	&	30	&	10	&	0	&	0	&	1	&	--	&	0.004	&	0.034	&	0.000	\\
run08	&	1D	&	RCP	&	3	&	10	&	3	&	0	&	1	&	Yes	&	0.645	&	0.212	&	0.016	\\
run09	&	1D	&	RCP	&	10	&	10	&	10	&	0	&	1	&	Yes	&	0.550	&	0.303	&	0.033	\\
run10	&	1D	&	RCP	&	100	&	10	&	100	&	0	&	1	&	Yes	&	0.474	&	0.272	&	0.428	\\
run11	&	1D	&	RCP	&	3	&	10	&	0	&	0	&	1	&	Yes	&	0.045	&	0.029	&	0.012	\\
run12	&	1D	&	RCP	&	10	&	10	&	0	&	0	&	1	&	Yes	&	0.016	&	0.104	&	0.006	\\
run13	&	1D	&	RCP	&	100	&	10	&	0	&	0	&	1	&	Yes	&	0.003	&	0.105	&	0.013	\\
run14	&	1D	&	RCP	&	30	&	30	&	30	&	0	&	1	&	Yes	&	0.437	&	0.387	&	0.254	\\
run15	&	1D	&	RCP	&	30	&	100	&	30	&	0	&	1	&	Yes	&	0.453	&	0.477	&	0.727	\\
run16	&	1D	&	RCP	&	30	&	30	&	0	&	0	&	1	&	Yes	&	0.002	&	0.045	&	0.005	\\
run17	&	1D	&	RCP	&	30	&	100	&	0	&	0	&	1	&	Yes	&	0.001	&	0.083	&	0.004	\\
run18	&	1D	&	RCP	&	30	&	10	&	30	&	0	&	0.3	&	Yes	&	0.483	&	0.318	&	0.457	\\
run19	&	1D	&	RCP	&	30	&	10	&	30	&	0	&	3	&	Yes	&	0.488	&	0.292	&	0.029	\\
run20	&	1D	&	RCP	&	30	&	10	&	0	&	0	&	0.3	&	Yes	&	0.007	&	0.084	&	0.026	\\
run21	&	1D	&	RCP	&	30	&	10	&	0	&	0	&	3	&	Yes	&	0.016	&	0.075	&	0.002	\\
run22	&	1D	&	RCP	&	30	&	10	&	30	&	5	&	1	&	Yes	&	0.493	&	0.318	&	0.108	\\
run23	&	1D	&	RCP	&	30	&	10	&	30	&	10	&	1	&	Yes	&	0.536	&	0.321	&	0.132	\\
run24	&	1D	&	RCP	&	30	&	10	&	30	&	15	&	1	&	Yes	&	0.490	&	0.328	&	0.144	\\
run25	&	1D	&	RCP	&	30	&	10	&	30	&	30	&	1	&	Yes	&	0.494	&	0.225	&	0.148	\\
run26	&	1D	&	RCP	&	30	&	10	&	30	&	45	&	1	&	Yes	&	0.460	&	0.201	&	0.178	\\ 
\hline \hline
\end{tabular}
\end{table*}

\begin{table*}
\caption{Electron features in the simulated runs. The model parameters of each run are listed in Table~\ref{tab1}.
\label{tab2}}
  \begin{tabular}{@{\extracolsep{10pt}}lccccccc}
    \hline \hline
Label & $\epsilon_{e,\max}$ & 
$\langle \epsilon_{e,{\rm total}} \rangle$ & 
$\langle \epsilon_{e,{\rm bulk}} \rangle$ & 
$\langle \epsilon_{e,{\rm hot}} \rangle$ & 
\multicolumn{3}{c}{$P \{ \epsilon_e > \epsilon_{\rm thr} \}$}
\\ \cline{6-8}
& [eV] & [eV] & [eV] & [eV] & 1 MeV & 10 MeV & 100 MeV \\ \hline
run01	& $	4.95 \times 10^8	$ & $	3.87 \times 10^6	$ & $	1.27 \times 10^5	$ & $	3.57 \times 10^7	$ & $	1.04 \times 10^{-1}	$ & $	9.17 \times 10^{-2}	$ & $	2.43 \times 10^{-3}	$ \\	
run02	& $	2.97 \times 10^6	$ & $	1.20 \times 10^4	$ & $	1.18 \times 10^4	$ & $	5.58 \times 10^5	$ & $	4.77 \times 10^{-5}	$ & --  & --  \\	
run03	& $	3.64 \times 10^8	$ & $	1.41 \times 10^6	$ & $	4.67 \times 10^4	$ & $	4.10 \times 10^7	$ & $	3.26 \times 10^{-2}	$ & $	2.85 \times 10^{-2}	$ & $	6.24 \times 10^{-4}	$ \\	
run04	& $	1.64 \times 10^8	$ & $	3.09 \times 10^6	$ & $	8.25 \times 10^4	$ & $	3.45 \times 10^7	$ & $	8.70 \times 10^{-2}	$ & $	7.95 \times 10^{-2}	$ & $	1.02 \times 10^{-3}	$ \\	
run05	& $	9.42 \times 10^7	$ & $	7.24 \times 10^4	$ & $	2.92 \times 10^4	$ & $	6.50 \times 10^6	$ & $	4.84 \times 10^{-3}	$ & $	1.43 \times 10^{-3}	$ & --  \\	
run06	& $	8.82 \times 10^7	$ & $	1.58 \times 10^5	$ & $	4.44 \times 10^4	$ & $	2.58 \times 10^6	$ & $	2.65 \times 10^{-2}	$ & $	2.17 \times 10^{-3}	$ & --  \\	
run07	& $	5.11 \times 10^5	$ & $	3.62 \times 10^4	$ & $	3.62 \times 10^4	$ & $	4.37 \times 10^5	$ & --  & --  & --  \\	
run08	& $	3.32 \times 10^7	$ & $	5.12 \times 10^4	$ & $	5.19 \times 10^3	$ & $	2.93 \times 10^6	$ & $	1.20 \times 10^{-2}	$ & $	2.24 \times 10^{-4}	$ & --  \\	
run09	& $	1.21 \times 10^8	$ & $	4.81 \times 10^5
$ & $	4.03 \times 10^4	$ & $	1.35 \times 10^7	$ & $
3.24 \times 10^{-2}	$ & $	2.19 \times 10^{-2}	$ & 
--  \\	
run10	& $	1.27 \times 10^9	$ & $	4.06 \times 10^7	$ & $	5.41 \times 10^5	$ & $	9.88 \times 10^7	$ & $	5.07 \times 10^{-1}	$ & $	3.36 \times 10^{-1}	$ & $	1.95 \times 10^{-1}	$ \\	
run11	& $	7.06 \times 10^6	$ & $	3.61 \times 10^3	$ & $	1.74 \times 10^3	$ & $	1.54 \times 10^5	$ & $	4.52 \times 10^{-4}	$ & --  & --  \\	
run12	& $	2.10 \times 10^7	$ & $	1.40 \times 10^4	$ & $	8.73 \times 10^3	$ & $	8.51 \times 10^5	$ & $	1.19 \times 10^{-3}	$ & $	3.53 \times 10^{-5}	$ & --  \\	
run13	& $	1.99 \times 10^8	$ & $	2.09 \times 10^5	$ & $	1.41 \times 10^5	$ & $	5.38 \times 10^6	$ & $	6.89 \times 10^{-3}	$ & $	1.33 \times 10^{-3}	$ & $	1.59 \times 10^{-4}	$ \\	
run14	& $	6.11 \times 10^8	$ & $	1.04 \times 10^7	$ & $	1.73 \times 10^5	$ & $	4.09 \times 10^7	$ & $	2.52 \times 10^{-1}	$ & $	2.35 \times 10^{-1}	$ & $	1.05 \times 10^{-2}	$ \\	
run15	& $	4.96 \times 10^8	$ & $	3.62 \times 10^7	$ & $	1.37 \times 10^6	$ & $	5.05 \times 10^7	$ & $	8.47 \times 10^{-1}	$ & $	6.64 \times 10^{-1}	$ & $	6.64 \times 10^{-2}	$ \\	
run16	& $	4.35 \times 10^7	$ & $	4.14 \times 10^4	$ & $	3.45 \times 10^4	$ & $	1.53 \times 10^6	$ & $	1.74 \times 10^{-3}	$ & $	9.61 \times 10^{-5}	$ & --  \\	
run17	& $	1.28 \times 10^8	$ & $	1.16 \times 10^5
$ & $	1.04 \times 10^5	$ & $	2.55 \times 10^6	$ & $
3.39 \times 10^{-3}	$ & $	2.20 \times 10^{-4}	$ & 
--  \\	
run18	& $	3.98 \times 10^8	$ & $	1.23 \times 10^7	$ & $	1.56 \times 10^5	$ & $	2.74 \times 10^7	$ & $	4.17 \times 10^{-1}	$ & $	3.17 \times 10^{-1}	$ & $	1.34 \times 10^{-2}	$ \\	
run19	& $	9.87 \times 10^8	$ & $	1.30 \times 10^6	$ & $	1.15 \times 10^5	$ & $	4.08 \times 10^7	$ & $	2.91 \times 10^{-2}	$ & $	2.72 \times 10^{-2}	$ & $	1.02 \times 10^{-3}	$ \\	
run20	& $	8.60 \times 10^7	$ & $	1.63 \times 10^5	$ & $	2.55 \times 10^4	$ & $	5.20 \times 10^6	$ & $	1.48 \times 10^{-2}	$ & $	4.46 \times 10^{-3}	$ & --  \\	
run21	& $	1.05 \times 10^8	$ & $	4.13 \times 10^4
$ & $	2.41 \times 10^4	$ & $	8.20 \times 10^6	$ & $
1.68 \times 10^{-3}	$ & $	6.08 \times 10^{-4}	$ & 
--  \\	
run22	& $	2.80 \times 10^8	$ & $	3.90 \times 10^6	$ & $	1.02 \times 10^5	$ & $	3.54 \times 10^7	$ & $	1.07 \times 10^{-1}	$ & $	9.83 \times 10^{-2}	$ & $	1.75 \times 10^{-3}	$ \\	
run23	& $	3.25 \times 10^8	$ & $	4.26 \times 10^6	$ & $	1.26 \times 10^5	$ & $	3.17 \times 10^7	$ & $	1.29 \times 10^{-1}	$ & $	1.13 \times 10^{-1}	$ & $	2.44 \times 10^{-3}	$ \\	
run24	& $	3.58 \times 10^8	$ & $	3.90 \times 10^6	$ & $	1.86 \times 10^5	$ & $	2.63 \times 10^7	$ & $	1.42 \times 10^{-1}	$ & $	1.10 \times 10^{-1}	$ & $	2.99 \times 10^{-3}	$ \\	
run25	& $	4.09 \times 10^8	$ & $	3.94 \times 10^6	$ & $	3.34 \times 10^5	$ & $	2.52 \times 10^7	$ & $	1.85 \times 10^{-1}	$ & $	9.87 \times 10^{-2}	$ & $	4.47 \times 10^{-3}	$ \\	
run26	& $	4.78 \times 10^8	$ & $	3.67 \times 10^6	$ & $	5.46 \times 10^5	$ & $	1.87 \times 10^7	$ & $	3.10 \times 10^{-1}	$ & $	8.93 \times 10^{-2}	$ & $	5.17 \times 10^{-3}	$ \\ 
    \hline \hline
  \end{tabular}
\end{table*}

\begin{table*}
\caption{Carbon-ion features in the simulated runs. The model parameters of each run are listed in Table~\ref{tab1}.
\label{tab3}}
  \begin{tabular}{@{\extracolsep{10pt}}lccccc}
    \hline \hline
Label & $\epsilon_{i,\max}$ & 
$\langle \epsilon_{i,{\rm total}} \rangle$ & 
\multicolumn{3}{c}{$P \{ \epsilon_i > \epsilon_{\rm thr} \}$}
\\ \cline{4-6}
& [eV/u] & [eV/u] & 1 MeV/u & 10 MeV/u & 100 MeV/u \\ \hline
run01	& $	1.75 \times 10^8	$ & $	1.17 \times 10^6	$ & $	7.72 \times 10^{-2}	$ & $	2.89 \times 10^{-2}	$ & $	7.18 \times 10^{-4}	$ \\	
run02	& $	3.58 \times 10^6	$ & $	1.27 \times 10^5	$ & $	2.53 \times 10^{-3}	$ & --  & --  \\	
run03	& $	1.41 \times 10^8	$ & $	5.71 \times 10^5	$ & $	2.87 \times 10^{-2}	$ & $	1.12 \times 10^{-2}	$ & $	4.13 \times 10^{-4}	$ \\	
run04	& $	8.47 \times 10^7	$ & $	8.75 \times 10^5	$ & $	6.26 \times 10^{-2}	$ & $	2.53 \times 10^{-2}	$ & --  \\	
run05	& $	9.45 \times 10^7	$ & $	3.25 \times 10^5	$ & $	2.59 \times 10^{-2}	$ & $	9.82 \times 10^{-3}	$ & --  \\	
run06	& $	9.87 \times 10^7	$ & $	5.21 \times 10^5	$ & $	4.39 \times 10^{-2}	$ & $	1.41 \times 10^{-2}	$ & --  \\	
run07	& $	2.03 \times 10^6	$ & $	1.39 \times 10^5	$ & $	1.29 \times 10^{-2}	$ & --  & --  \\	
run08	& $	1.14 \times 10^7	$ & $	8.46 \times 10^3	$ & $	2.18 \times 10^{-3}	$ & $	4.92 \times 10^{-5}	$ & --  \\	
run09	& $	5.51 \times 10^7	$ & $	1.34 \times 10^5	$ & $	1.87 \times 10^{-2}	$ & $	2.91 \times 10^{-3}	$ & --  \\	
run10	& $	5.74 \times 10^8	$ & $	1.21 \times 10^7	$ & $	7.45 \times 10^{-1}	$ & $	1.95 \times 10^{-1}	$ & $	2.45 \times 10^{-2}	$ \\	
run11	& $	1.71 \times 10^6	$ & $	1.18 \times 10^3	$ & $	6.89 \times 10^{-5}	$ & --  & --  \\	
run12	& $	2.40 \times 10^7	$ & $	4.62 \times 10^4	$ & $	7.78 \times 10^{-3}	$ & $	6.71 \times 10^{-4}	$ & --  \\	
run13	& $	9.35 \times 10^7	$ & $	4.68 \times 10^6	$ & $	8.26 \times 10^{-1}	$ & $	4.42 \times 10^{-2}	$ & --  \\	
run14	& $	2.96 \times 10^8	$ & $	4.63 \times 10^6	$ & $	2.71 \times 10^{-1}	$ & $	1.19 \times 10^{-1}	$ & $	3.15 \times 10^{-3}	$ \\	
run15	& $	5.21 \times 10^8	$ & $	1.90 \times 10^7	$ & $	6.75 \times 10^{-1}	$ & $	3.95 \times 10^{-1}	$ & $	3.33 \times 10^{-2}	$ \\	
run16	& $	5.29 \times 10^7	$ & $	5.45 \times 10^5	$ & $	1.11 \times 10^{-1}	$ & $	5.97 \times 10^{-3}	$ & --  \\	
run17	& $	1.44 \times 10^8	$ & $	3.33 \times 10^6	$ & $	9.49 \times 10^{-1}	$ & $	3.03 \times 10^{-3}	$ & $	6.59 \times 10^{-5}	$ \\	
run18	& $	1.81 \times 10^8	$ & $	4.08 \times 10^6	$ & $	3.39 \times 10^{-1}	$ & $	1.15 \times 10^{-1}	$ & $	2.46 \times 10^{-3}	$ \\	
run19	& $	1.40 \times 10^8	$ & $	3.93 \times 10^5	$ & $	2.23 \times 10^{-2}	$ & $	8.88 \times 10^{-3}	$ & $	1.19 \times 10^{-4}	$ \\	
run20	& $	9.71 \times 10^7	$ & $	1.07 \times 10^6	$ & $	1.24 \times 10^{-1}	$ & $	3.16 \times 10^{-2}	$ & --  \\	
run21	& $	7.72 \times 10^7	$ & $	1.01 \times 10^5	$ & $	5.94 \times 10^{-3}	$ & $	3.24 \times 10^{-3}	$ & --  \\	
run22	& $	1.74 \times 10^8	$ & $	1.27 \times 10^6	$ & $	8.42 \times 10^{-2}	$ & $	3.70 \times 10^{-2}	$ & $	7.77 \times 10^{-4}	$ \\	
run23	& $	1.55 \times 10^8	$ & $	1.28 \times 10^6	$ & $	9.28 \times 10^{-2}	$ & $	3.86 \times 10^{-2}	$ & $	6.39 \times 10^{-4}	$ \\	
run24	& $	1.32 \times 10^8	$ & $	1.31 \times 10^6	$ & $	1.01 \times 10^{-1}	$ & $	4.10 \times 10^{-2}	$ & $	3.64 \times 10^{-4}	$ \\	
run25	& $	8.58 \times 10^7	$ & $	8.99 \times 10^5	$ & $	9.55 \times 10^{-2}	$ & $	2.27 \times 10^{-2}	$ & --  \\	
run26	& $	6.36 \times 10^7	$ & $	8.02 \times 10^5	$ & $	1.11 \times 10^{-1}	$ & $	1.66 \times 10^{-2}	$ & --  \\ 
    \hline \hline
  \end{tabular}
\end{table*}

\section{Test particle trajectories in standing whistler waves \label{app2}}

The equations for the electron motion in CP standing waves are solved analytically to interpret the physical mechanism of electron acceleration.
Here, we assume the same $k_0$ and $\omega_0$ for the wavenumber and frequency of the counter-propagating electromagnetic waves that form the standing wave.
Suppose that the magnetic field $B_{\rm ext}$ is oriented in the $x$ direction of the Cartesian coordinate system.
The wave amplitudes propagating in the positive direction of $x$ are denoted by ($E_0^{+}$, $B_0^{+}$), and those in the reverse direction by ($E_0^{-}$, $B_0^{-}$).
In this case, the electromagnetic fields of the standing wave can be described as
\begin{eqnarray}
\bm{E} &=& \bm{E}_0^{+} + \bm{E}_0^{-} \nonumber \\
&=& - (E_0^{+}+E_0^{-}) \sin k_0 x  
\left(
\begin{array}{c}
0 \\
\cos \omega_0 t \\
\sin \omega_0 t 
\end{array}  \right) \nonumber \\
&& -  (E_0^{+}-E_0^{-}) \cos k_0 x  
\left(
\begin{array}{c}
0 \\
- \sin \omega_0 t \\
\cos \omega_0 t 
\end{array}  \right) \;,
\label{eq:esw}
\end{eqnarray}
\begin{eqnarray}
\bm{B} &=& \bm{B}_{\rm ext} + \bm{B}_0^{+} + \bm{B}_0^{-} \nonumber \\
&=& 
\left(
\begin{array}{c}
B_{\rm ext} \\
0 \\
0
\end{array}
\right) + (B_0^{+}+B_0^{-}) \cos k_0 x  
\left(
\begin{array}{c}
0 \\
\cos \omega_0 t \\
\sin \omega_0 t 
\end{array}  \right) \nonumber \\
&& + (B_0^{+}-B_0^{-}) \sin k_0 x  
\left(
\begin{array}{c}
0 \\
\sin \omega_0 t \\
- \cos \omega_0 t 
\end{array}  \right) \;.
\label{eq:bsw}
\end{eqnarray}
The amplitude of the electromagnetic fields is connected by the following relationship,
\begin{equation}
\bm{B}^{\pm} = \frac{N}{c} \bm{E}^{\pm} \;,
\end{equation}
where $N$ is the refractive index defined by Eq.~(\ref{eq:n}).
It should be noticed that the envelopes of the transverse components $E_{\perp}$ and $B_{\perp}$ are a function of position $x$ alone and independent of time $t$, 
\begin{equation}
E_{\perp}^2 = (E_0^{+} + E_0^{-})^2 \sin^2 k_0 x
+ (E_0^{+} - E_0^{-})^2 \cos^2 k_0 x \;,
\end{equation}
\begin{equation}
B_{\perp}^2 = (B_0^{+} + B_0^{-})^2 \cos^2 k_0 x
+ (B_0^{+} - B_0^{-})^2 \sin^2 k_0 x \;.
\end{equation}

The electron motion in the standing wave is solved by treating it as a test particle in the electromagnetic fields defined by Eqs.~(\ref{eq:esw}) and (\ref{eq:bsw}).
The basic equations to be solved are
\begin{equation}
\frac{d \bm{p}}{dt} = - e \left( \bm{E} + 
\frac{\bm{p}}{\gamma m_e} \times \bm{B} \right) \;,
\label{eq:dpdt}
\end{equation}
\begin{equation}
\frac{dx}{dt} = \frac{p_x}{\gamma m_e} \;,
\label{eq:dxdt}
\end{equation}
where the momentum $\bm{p} = \gamma m_e \bm{v}$ and the Lorentz factor $\gamma = [ 1 + (\bm{p}/m_e c)^2 ]^{1/2}$.

\begin{table*}
\caption{Conditions and momenta at various fixed points
\label{tab:fp}}
\begin{tabular}{@{\extracolsep{10pt}}lccc}
\hline \hline
Type & \multicolumn{2}{c}{Condtions for $\widehat{x}$ and $\psi$} & $\widetilde{p}_{\perp}$ 
\\ \hline
FP1 & $\cos \widehat{x} = \cos \psi = 0$ &
$\sin \widehat{x} \sin \psi = 1$ &
$\widetilde{p}_{\perp} ( 
{\widetilde{B}_{\rm ext}}/{\gamma} - 1 ) = ( 1 + R ) a_0$ \\
FP2 & $\cos \widehat{x} = \cos \psi = 0$ &
$\sin \widehat{x} \sin \psi = - 1$ &
$\widetilde{p}_{\perp} ( 
1 - {\widetilde{B}_{\rm ext}}/{\gamma} ) = ( 1 + R ) a_0$ \\
FP3 & $\sin \widehat{x} = \sin \psi = 0$ &
$\cos \widehat{x} \cos \psi = 1$ &
$\widetilde{p}_{\perp} ( 
1 - {\widetilde{B}_{\rm ext}}/{\gamma} ) = ( 1 - R ) a_0$ \\
FP4 & $\sin \widehat{x} = \sin \psi = 0$ &
$\cos \widehat{x} \cos \psi = - 1$ &
$\widetilde{p}_{\perp} ( 
{\widetilde{B}_{\rm ext}}/{\gamma} - 1 ) = ( 1 - R ) a_0$ \\
\hline \hline
\end{tabular}
\end{table*}

Using $\bm{p} = (p_{\parallel}, p_{\perp} \cos \phi, p_{\perp} \sin \phi)$ and $\psi = \phi - \omega_0 t$, Eqs.~(\ref{eq:dpdt}) and (\ref{eq:dxdt}) are rewritten as
\begin{equation}
\frac{d \widetilde{p}_{\parallel}}{d \widehat{t}} =
N a_0 \frac{\widetilde{p}_{\perp}}{\gamma}
\left[ (1+R) \cos \widehat{x} \sin \psi
+ (1-R) \sin \widehat{x} \cos \psi 
\right] \;,
\label{eq:ppara}
\end{equation}
\begin{eqnarray}
\frac{d \widetilde{p}_{\perp}}{d \widehat{t}} &=&
a_0
\left[ (1+R) \sin \widehat{x} \cos \psi
+ (1-R) \cos \widehat{x} \sin \psi 
\right] \nonumber \\
&-& N a_0 \frac{\widetilde{p}_{\parallel}}{\gamma}
\left[ (1+R) \cos \widehat{x} \sin \psi
+ (1-R) \sin \widehat{x} \cos \psi 
\right]  \;,
\label{eq:pperp}
\end{eqnarray}
\begin{eqnarray}
\frac{d \psi}{d \widehat{t}} &=&
a_0 \frac1{\widetilde{p}_{\perp}}
\left[ (1+R) \sin \widehat{x} \sin \psi
- (1-R) \cos \widehat{x} \cos \psi 
\right] \nonumber \\
&-& N a_0 \frac{\widetilde{p}_{\parallel}} {\gamma \widetilde{p}_{\perp}}
\left[ (1+R) \cos \widehat{x} \cos \psi
- (1-R) \sin \widehat{x} \sin \psi 
\right] \nonumber \\
&+& \frac{\widetilde{B}_{\rm ext}}{\gamma} - 1  \;,
\label{eq:psi}
\end{eqnarray}
\begin{equation}
\frac{d \widehat{x}}{d \widehat{t}} = 
N \frac{\widetilde{p}_{\parallel}}{\gamma}  \;,
\label{eq:x}
\end{equation}
where the momentum, position, and time are normalized by $\widetilde{p_*} = p_* / (m_e c)$, $\widehat{x} = k_0 x$, and $\widehat{t} = \omega_0 t$, respectively.
The field amplitudes of the waves are assumed to be $E^{+} = E_0$ and $E^{-} = R E^{+}$.
Then the energy evolution of the electron is determined through the equation below,
\begin{equation}
\frac{d \gamma}{d \widehat{t}} = 
a_0 \frac{\widetilde{p}_{\perp}}{\gamma}
\left[ (1+R) \sin \widehat{x} \cos \psi
+ (1-R) \cos \widehat{x} \sin \psi 
\right] \;.
\end{equation}

The fixed points, at which all the time derivatives of Eqs.~(\ref{eq:ppara})--(\ref{eq:x}) vanish, are helpful to understand the resonance characteristics of an electron.
By setting $p_{\parallel} = 0$, the requirement for a fixed point is derived as $\cos \widehat{x} = \cos \psi = 0$ or $\sin \widehat{x} = \sin \psi = 0$.
The momentum values at the fixed points in our situation are summarized in Table~\ref{tab:fp}.
The fixed points, FP1 and FP2, are the same as those already reported \cite{matsukiyo09}, but the other two have appeared as a result of considering two waves with different amplitudes.

As for the case of $\cos \widehat{x} = \cos \psi = 0$, let us consider an electron at $\widehat{x} = \pi / 2$, i.e., at the trough of the magnetic field.
Then, the condition for the FP1 is satisfied at $\psi = \pi / 2$.
In the non-relativistic limit ($\gamma \approx 1$), the momentum at FP1 is given by $\widetilde{p}_{\perp} \approx (1 + R) a_0 /(\widetilde{B}_{\rm ext} - 1)$.
Under a strong magnetic field ($\widetilde{B}_{\rm ext} > 1$), it becomes a stable fixed point \cite{matsukiyo09}.
In the relativistic limit ($\gamma \approx \widetilde{p}_{\perp}$), the momentum will become $\widetilde{p}_{\perp} \approx \widetilde{B}_{\rm ext} - (1+R) a_0$.
Although an extra condition $\widetilde{B}_{\rm ext} \gg (1+R) a_0$ is needed for it to be valid, this solution is known to be unstable \cite{matsukiyo09}.
On the other hand, at $\psi = 3 \pi / 2$, we have the stable FP2, which is given by $\widetilde{p}_{\perp} \approx (1+R) a_0 + \widetilde{B}_{\rm ext}$ in the relativistic limit.
The relativistic FP2 is related to the cyclotron resonance, and its existence is the key to electron acceleration.
The momentum at FP2 in the small-amplitude limit, $\widetilde{p}_{\perp} = (\widetilde{B}_{\rm ext}^2 - 1)^{1/2}$, satisfies the resonance conditions for both of the two counter waves simultaneously \cite{matsukiyo09,lee13}.

The relativistic momentum at the FP3, $\widetilde{p}_{\perp} \approx (1-R) a_0 + \widetilde{B}_{\rm ext}$, is always smaller than that of the FP2.
Therefore, for the discussion of the maximum momentum, we should focus on the electron motions at $\widehat{x} = \pi / 2$.

Assuming $\widetilde{p}_{\parallel} = 0$, the equations of motion for an electron at $\widehat{x} = \pi / 2$ are given by
\begin{equation}
\frac{d \widetilde{p}_{\parallel}}{d \widehat{t}} =
N (1-R) a_0 \frac{\widetilde{p}_{\perp}}{\gamma} 
\cos \psi \;,
\label{eom1}
\end{equation}
\begin{equation}
\frac{d \widetilde{p}_{\perp}}{d \widehat{t}} =
(1+R) a_0 
\cos \psi \;,
\label{eom2}
\end{equation}
\begin{equation}
\frac{d \psi}{d \widehat{t}} = 
- (1+R) a_0  
\frac1{\widetilde{p}_{\perp}}
\sin \psi 
+ \frac{\widetilde{B}_{\rm ext}}{\gamma} - 1  \;.
\label{eom3}
\end{equation}
If the amplitude of the opposing wave is the same ($R=1$), there is no force in the parallel direction.
It is because the perpendicular component of the magnetic field is zero at the trough of the magnetic field. 
Thanks to this property, the governing equations for the electron gyration motion are closed only with Eqs.~(\ref{eom2}) and (\ref{eom3}).
However, if the amplitudes of the counter waves are different, the magnetic field of the standing wave is finite everywhere.
Then, strictly speaking, the parallel motion is coupled with the transverse momentum.
In the non-relativistic limit of $\widetilde{p}_{\perp}$, the time derivative of $\widetilde{p}_{\parallel}$ is negligible compared to that of $\widetilde{p}_{\perp}$ by a factor of $\widetilde{p}_{\perp} / \gamma \ll 1$.
On the other hand, for the case of relativistic momentum, these time derivatives are in the same order, so we have to solve the equations of motion in the horizontal direction at the same time.

For simplicity, we will consider only Eqs. (\ref{eom2}) and (\ref{eom3}) in the analysis of this paper, and define the Hamiltonian to describe the electron motions following the previous works \cite{matsukiyo09,isayama23}.
Note that for the case of $R=1$, our analysis can be regarded as exact, while it corresponds to approximated solutions for the non-relativistic momentum if $R \neq 1$.
The equations of motion~(\ref{eom2}) and (\ref{eom3}) are combined into the following equation,
\begin{equation}
H (\chi, \psi) = A \chi^{1/2} \sin \psi - B (\chi + 1)^{1/2} + \chi \;,
\label{eq:h}
\end{equation}
where $\chi = \widetilde{p}_{\perp}^2$, $A = 2 (1+R) a_0$, and $B = 2 \widetilde{B}_{\rm ext}$.
Then, Eq.~(\ref{eq:h}) satisfies the relations $d \chi / d \widehat{t} = \partial H / \partial \psi$ and $d \psi / d \widehat{t} = - \partial H / \partial \chi$ simaltaneously.
The electron trajectories in the $\chi$-$\psi$ coordinate can be obtained as constant $H$ contour lines.
This equation~(\ref{eq:h}) gives the basis of the analysis discussed in Sec.~\ref{sec3d}.
This relation is valid at the trough (crest) of the magnetic (electric) field in the standing wave, that is $\widehat{x} = ( 2 n - 1 )  \pi / 2$ where $n$ is an integer. 
These locations correspond to $\widetilde{x} \approx ( 2 n - 1 ) / 4$ in our simulations.



%



\end{document}